\documentclass[envcountsect,envcountsame,11pt]{llncs}
\usepackage[a4paper,hmargin=1.2in,vmargin=1.25in]{geometry}

\usepackage{latexsym}
\usepackage{amssymb}
\usepackage{amsmath}
\usepackage{float}
\usepackage{framed}
\usepackage{qip}
\usepackage{bbm}  
\usepackage{tabularx}
\usepackage{xspace}
\usepackage{enumerate}
\usepackage{graphicx}

\usepackage{epsfig}
\usepackage{wrapfig}

\usepackage{color}

\usepackage{ifpdf}

\allowdisplaybreaks
\ifpdf

\else

\fi

\ifpdf
\newcommand{\PDForPSinclude}[2]{{ \includegraphics[width=#2]{#1.pdf} }}
\else
\newcommand{\PDForPSinclude}[2]{{ \includegraphics[width=#2]{#1.eps} }}
\fi

\newcommand{\epsclose}[1][\eps]{\approx_{#1}}

\newcommand{\close}[1]{\approx_{#1}}

\newcommand*{\F}{\mathcal{F}}

\newcommand*{\X}{\mathcal{X}}
\newcommand*{\cY}{\mathcal{Y}}

\newcommand{\set}[1]{\{#1\}}
\newcommand{\Set}[2]{\{ #1 : #2\}}

\newcommand{\A}{{\sf A}}
\newcommand{\dA}{{\sf A}'}
\newcommand{\B}{{\sf B}}
\newcommand{\dB}{{\sf B}'}
\newcommand{\hA}{\hat{\sf A}}
\newcommand{\dhA}{\hat{\sf A}'}
\newcommand{\hB}{\hat{\sf B}}
\newcommand{\dhB}{\hat{\sf B}'}

\newcommand{\onetwo}[1][2]{\mbox{\textsl{1\hspace{-0.1ex}-#1}}}
\def\-{\hspace{-0.1ex}-\hspace{-0.1ex}}           

\newcommand{\Rand}{\textsl{Rand}}                 
\newcommand{\lStringOT}[1][2]{\textsl{\onetwo[#1]\:OT\,$^{\ell}$}\xspace}    
\newcommand{\RandlStringOT}[1][2]{\Rand\:\lStringOT[#1]}    

\newcommand{\ID}{{\sf \textsl{ID}}\xspace}

\newcommand{\eps}{\varepsilon}

\newcommand{\nord}{{}}

\newcommand{\MC}[3]{#1 \leftrightarrow #2 \leftrightarrow #3}

\newcommand*{\assign}{\ensuremath{\kern.5ex\raisebox{.1ex}{\mbox{\rm:}}\kern -.3em =}}

\newcommand{\ev}{{\cal E}}
\renewcommand{\and}{and}
\renewcommand*{\id}{\mathrm{id}}
\renewcommand*{\I}{\mathbbm{I}}

\newcommand{\etal}{{\it et al.}}

\newcommand{\qp}{\pi}
\newcommand{\cp}{\Sigma}

\newcommand{\emp}{\emptyset}

\newcommand{\eqq}{\stackrel{\text{\tiny?}}{=}}

\newcommand{\ol}[1]{\overline{#1}}

\newcommand{\myparagraph}[1]{\medskip\noindent{\sc #1}}

\newcommand{\delete}[1]{}
\newcommand{\remove}[1]{}

\title{Composing Quantum Protocols in a Classical Environment}

\author{Serge Fehr \and Christian Schaffner}

\institute{ Centrum voor Wiskunde en Informatica (CWI) \\
    Amsterdam, The Netherlands \\ \email{\set{S.Fehr,C.Schaffner}@cwi.nl}}

\date{\today}

\begin{document}

\maketitle

\begin{abstract}
  We propose a general security definition for cryptographic quantum
  protocols that implement classical non-reactive two-party tasks. The
  definition is expressed in terms of simple
  quantum-information-theoretic conditions which must be satisfied by
  the protocol to be secure. The conditions are uniquely determined by
  the ideal functionality $\F$ defining the cryptographic task to be
  implemented.  We then show the following composition result. If
  quantum protocols $\qp_1,\ldots,\qp_\ell$ securely implement ideal
  functionalities $\F_1,\ldots,\F_\ell$ according to our security
  definition, then any purely {\em classical} two-party protocol,
  which makes sequential calls to $\F_1,\ldots,\F_\ell$, is equally
  secure as the protocol obtained by replacing the calls to
  $\F_1,\ldots,\F_\ell$ with the respective quantum protocols
  $\qp_1,\ldots,\qp_\ell$.  Hence, our approach yields the minimal
  security requirements which are strong enough for the typical use of
  quantum protocols as subroutines within larger classical
  schemes. Finally, we show that recently proposed quantum protocols
  for oblivious transfer and secure identification in the
  bounded-quantum-storage model satisfy our security definition, and
  thus compose in the above sense.
\\[1ex]
\end{abstract}

\section{Introduction}

\vspace{-1ex}\subsubsection{Background.}

Finding the right security definition for a cryptographic task is a
non-trivial fundamental question in cryptography. From a theoretical
point of view, one would like definitions to be as strong as possible
in order to obtain strong composability guarantees. However, this often
leads to impossibility results or to very complex and inefficient
schemes. Therefore, from a practical point of view, one may also
consider milder security definitions which allow for efficient
schemes, but still offer ``good enough'' security.

It is fair to say that in computational cryptography, the question of
defining security and the trade-offs that come along with these
definitions are by now quite well understood. The situation is
different in quantum cryptography. For instance, it was realized only
recently that the standard security definition of quantum
key-agreement does not guarantee the desired kind of security and some
work was required to establish the right security
definition~\cite{GotLo03,RK05,BHLMO05,Renner05,KRBM07}. In~\cite{BM04,Unruh04},
strong security definitions for general quantum protocols were
proposed by translating Canetti's universal-composability framework
and Backes, Pfitzmann and Waidner's reactive-simulatability model,
respectively, into the quantum setting. The resulting security
definitions are very strong and guarantee full composability. 
However, they are complex and hard to achieve. Indeed, so far they have been actually used and shown to be achievable only in a couple of isolated cases: quantum key distribution~\cite{BHLMO05} and quantum multi-party computation with dishonest minority~\cite{BCGHS05}.
It is still
common practice in quantum cryptography that every paper proposes its
own security definition of a certain task and proves security
with respect to the proposed definition. However, it usually remains unclear
whether these definitions are strong enough to guarantee any kind
of composability, and thus whether protocols that meet the definition really behave as expected.

\vspace{-2ex}\subsubsection{Contribution.}

We propose a general security definition for quantum protocols that
implement cryptographic two-party tasks. The definition is in
terms of simple quantum-information-theoretic security conditions that
must be satisfied for the protocol to be secure. In particular,
the definition does not involve additional entities like a ``simulator'' or an
``environment''. The security conditions are uniquely determined by
the {\em ideal functionality} that defines the cryptographic task to
be realized.  Our definition applies to any {\em non-reactive,
  classical} ideal functionality~$\F$, which obtains classical (in the
sense of non-quantum) input from the two parties, processes the
provided input according to its specification, and outputs the
resulting classical result to the parties. A typical example for such
a functionality/task is oblivious transfer (OT).  Reactive
functionalities, i.e.~functionalities that have several phases
(like e.g.\ bit commitment), or functionalities that take quantum
input and/or produce quantum output are not the scope of this
paper.

We show the following composition result. If quantum protocols
$\qp_1,\ldots,\qp_\ell$ securely implement ideal functionalities
$\F_1,\ldots,\F_\ell$ according to our security definition, then any
purely {\em classical} two-party protocol, which makes sequential
calls to $\F_1,\ldots,\F_\ell$, is equally secure as the protocol
obtained by replacing the calls to $\F_1,\ldots,\F_\ell$ with the
respective quantum subroutines $\qp_1,\ldots,\qp_\ell$.  We stress
that our composition theorem, respectively our security definition,
only allows for the composition of quantum sub-protocols into a {\em
  classical} outer protocol. This is a trade-off which allows for
milder security definitions (which in turn allows for simpler and more
efficient implementations) but still offers security in realistic
situations. Indeed, current technology is far from being able to
execute quantum algorithms or protocols which involve complicated
quantum operations and/or need to keep a quantum state ``alive'' for
more than a tiny fraction of a second. Thus, the best one can hope for
in the near future in terms of practical quantum algorithms is that
certain small subroutines, like key-distribution or OT, may be
implemented by quantum protocols, while the more complex outer
protocol remains classical. 
From a more theoretical point of view, our general security definition
expresses what security properties a quantum protocol must satisfy in
order to be able to instantiate a basic cryptographic primitive
upon which an information-theoretic cryptographic construction is
based. For instance, it expresses the security properties a quantum
OT\footnote{We are well aware that quantum OT is impossible without
  any restriction on the adversary, but it becomes possible
  for instance when restricting the adversary's quantum
  memory~\cite{DFSS05,DFRSS07}. } needs to satisfy so that Kilian's
classical\footnote{Here, ``classical'' can be understood as
  ``non-quantum'' as well as ``being a classic''.} construction of
general secure function evaluation based on OT~\cite{Kilian88} remains
secure when instantiating the OT primitive by a quantum protocol. 
Alternatively, our security conditions can also be viewed as providing the {\em minimal} requirements for a quantum protocol to behave as expected. 

Finally, we show that the \emph{ad-hoc} security definitions proposed
by Damg{\aa}rd, Fehr, Salvail and Schaffner for their \mbox{1-2 OT}
and secure-identification protocols in the bounded-quantum-storage
model \cite{DFRSS07,DFSS07} imply (and are likely to be equivalent) to
the corresponding security definitions obtained from our
approach.\footnote{Interestingly, this is not true for the definition
  of Rabin OT given in the first paper in this line of
  research~\cite{DFSS05}, and indeed in the full version of that
  paper, it is mentioned that their definition poses some
  ``composability problems'' (this problem though has been fixed in
  the journal version~\cite{DFSS08}). This supports our claim that
  failure of satisfying our security definition is strong evidence for
  a security problem of a quantum protocol. }  This implies
composability in the above sense for these quantum protocols in the
bounded-quantum-storage model.

\vspace{-2ex}\subsubsection{Related work.}

In the classical setting, Cr{\'e}peau, Savvides, Schaffner and
Wullschleger proposed information-theoretic conditions for two-party
secure function evaluation~\cite{CSSW06}, though restricted to the
{\em perfect} case, where the protocol is not allowed to make any
error.  They show equivalence to a simulation-based definition that
corresponds to the standard framework of
Goldreich~\cite{Goldreich04}. Similar conditions have been
subsequently found~by Cr\'epeau and Wullschleger for the case of
non-perfect classical protocols~\cite{CW08}. Our work can be seen as
an extension of \cite{CSSW06,CW08} to the setting where classical
subroutines are implemented by quantum protocols.

As pointed out and discussed above, general frameworks for universal composability in the quantum setting have been established in~\cite{BM04,Unruh04}. 
The composability of protocols in the bounded-quantum-storage model has
recently been investigated by Wehner and
Wullschleger~\cite{WW08}. They propose security definitions that
guarantee sequential composability of quantum protocols within {\em
  quantum} protocols. This is clearly a stronger composition result
than we obtain (though restricted to the bounded-quantum-storage
model) but comes at the price of a more demanding security
definition. And indeed, whereas we show that the simple definitions
used in \cite{DFSS05,DFRSS07} already guarantee composability into
classical protocols without any modifications to the original
parameters and proofs, \cite{WW08} need to strengthen the
quantum-memory bound (and re-do the security proof) in order to show
that the 1-2 OT protocol from~\cite{DFRSS07} meets their strong
security definition. As we argued above, this is an overkill in many
situations.

\section{Notation} \label{sec:notation}
\myparagraph{Quantum States.} We assume the reader's familiarity with
basic notation and concepts of quantum information processing~\cite{NC00}.

Given a bipartite quantum state $\rho_{XE}$, we say that $X$ is {\em classical} if $\rho_{XE}$ is of the form $\rho_{XE} = \sum_{x \in \X} P_X(x) \proj{x} \otimes \rho_E^x$ for a
probability distribution $P_X$ over a finite set $\X$. This can be understood in that the state of
the quantum register $E$ depends on the classical random variable $X$, in the sense that $E$ is in state $\rho_E^x$ exactly if $X = x$. 
For any event ${\cal E}$ defined by $P_{\ev|X}(x) = P[\ev|X\!=\!x]$ for all $x$, we may then write 
\begin{equation}\label{eq:states}
\rho_{XE|{\cal E}} \assign \sum_x P_{X|{\cal E}}(x) \proj{x} \otimes \rho_E^x \enspace .
\end{equation}
When we omit
registers, we mean the partial trace over these register, for instance $\rho_{E|{\cal E}}
= \trace[X]{\rho_{XE|{\cal E}}} = \sum_x P_{X|{\cal E}}(x) \rho_E^x$, which describes $E$ given that the event $\cal E$ occurs.

This notation extends naturally to states that depend on
several classical random variables $X$, $Y$ etc., defining the density matrices
$\rho_{XYE}$, $\rho_{XYE|\ev}$, $\rho_{YE|X=x}$ etc.  We
tend to slightly abuse notation and write $\rho^x_{YE} =
\rho_{XE|X=x}$ and $\rho^x_{YE|\ev} = \rho_{YE|X=x,\ev}$,
as well as $\rho^x_{E} = \tr_Y(\rho^x_{YE})$ and
$\rho^x_{E|\ev} = \tr_Y(\rho^x_{YE|\ev})$.
Given a state $\rho_{XE}$ with classical $X$, by saying
that ``there exists a classical random variable $Y$ such that $\rho_{XYE}$
satisfies some condition'', we mean that $\rho_{XE}$ can be
understood as $\rho_{XE} = \tr_Y(\rho_{XYE})$ for some state
$\rho_{XYE}$ with classical $X$ and $Y$, and that $\rho_{XYE}$
satisfies the required condition.\footnote{This is similar to the case
  of distributions of classical random variables where given $X$ the
  existence of a certain $Y$ is understood that there exists a certain
  joint distribution $P_{XY}$ with $\sum_y P_{XY}(\cdot,y) = P_X$. }

$X$ is independent of $E$ (in that $\rho^x_E$ does not depend on $x$) if and only if $\rho_{XE} = \rho_X \otimes \rho_E$, which in particular implies
that no information on $X$ can be learned by observing only $E$. 
Similarly, $X$ is random and independent of $E$ if and only if $\rho_{XE} = \frac{1}{|{\cal X}|}\I \otimes \rho_E$, where $\frac{1}{|{\cal X}|}\I$ is the density matrix of the fully mixed state of suitable dimension. 

We also need to express that a random variable $X$ is independent of a quantum state 
$E$ {\em when given a random variable $Y$}. This means that when given $Y$, the state $E$ gives no additional information on $X$. 
Yet another way to understand this is that $E$ is obtained from $X$ and $Y$ by solely processing $Y$. 
Formally, adopting the notion introduced in~\cite{DFSS07}, this is expressed by requiring that $\rho_{X Y E}$ equals $\rho_{X\leftrightarrow Y \leftrightarrow E}$, where the latter is defined as
$$
\rho_{X\leftrightarrow Y \leftrightarrow E} :=  \sum_{x,y}P_{X Y}(x,y)\proj{x} \otimes \proj{y} \otimes \rho_{E}^y \, .
$$ 
In other words, $\rho_{X Y E} = \rho_{X\leftrightarrow Y \leftrightarrow E}$ precisely if $\rho_E^{x,y} = \rho_E^{y}$ for all $x$ and $y$. 
This notation naturally extends to 
$\rho_{X\leftrightarrow Y \leftrightarrow E|{\cal E}} = \sum_{x,y}P_{X Y|\ev}(x,y)\proj{x} \otimes \proj{y} \otimes \rho_{E|\ev}^{y}$.
 
Full (conditional) independence is often too strong a requirement, and
it usually suffices to be ``close'' to such a situation.  Closeness of
two states $\rho$ and $\sigma$ is measured in terms of their trace
distance $\delta(\rho,\sigma) = \frac{1}{2} \tr(|\rho-\sigma|)$, where
for any operator $A$, $|A|$ is defined as \smash{$|A| \assign \sqrt{A
    A^\dag}$}. We write $\rho \epsclose \sigma$ to denote that
$\delta(\rho,\sigma) \leq \eps$, and we then say that $\rho$ and
$\sigma$ are $\eps$-close.  It is known that $\eps$-closeness is
preserved under any quantum operation; this in particular
implies that if $\rho \epsclose \sigma$ then no observer can
distinguish $\rho$ from $\sigma$ with advantage greater
than~$\eps$~\cite{RK05}.  For states $\rho_{XE}$ and $\rho_{X'E'}$
with classical $X$ and $X'$, it is not hard to see that
$\delta(\rho_{XE}, \rho_{X'E'}) = \sum_x \delta(P_X(x) \rho_E^x, P_{X'}(x) \rho_{E'}^x )$, and thus $\delta(\rho_{XE},
\rho_{X'E'}) = \sum_x P_X(x) \delta(\rho_E^x, \rho_{E'}^x)$ if $P_X =
P_{X'}$.  In case of purely classical states $\rho_X$ and $\rho_{X'}$,
the trace distance coincides with the statistical distance of the
random variables $X$ and $X'$: $\delta(\rho_X,\rho_{X'}) = \frac12
\sum_x |P_X(x)-P_{X'}(x)|$, and we then write $P_{X} \epsclose
P_{X'}$, or $X \epsclose X'$, instead of $\rho_X \epsclose \rho_{X'}$.

We will make use of the following lemmas whose proofs are given in Appendix~\ref{app:proofs}.

\begin{lemma}\label{lemma:right->mid}
\begin{enumerate}
\item If $\rho_{XYZE} \epsclose \rho_{\MC{X}{Y}{ZE}}$ then $\rho_{XYZE}
\epsclose[2\eps] \rho_{\MC{X}{YZ}{E}}$. 
\item If $\rho_{XZE}
\epsclose \rho_X \otimes \rho_{ZE}$ then $\rho_{XZE} \epsclose[2\eps]
\rho_{\MC{X}{Z}{E}}$.
\item If $\rho_{XZE} \epsclose \I/|{\cal X}| \otimes
\rho_{ZE}$, then $\rho_{XZE} \epsclose[4\eps] \rho_{\MC{X}{Z}{E}}$.
\end{enumerate}
\end{lemma}

\begin{lemma}\label{lemma:extend}
If $\rho_{XYE} \epsclose \rho_{\MC{X}{Y}{E}}$ then $\rho_{Xf(X,Y)YE} \epsclose \rho_{\MC{Xf(X,Y)}{Y}{E}}$ for any function $f$. 
\end{lemma}

\begin{lemma} \label{lemma:event} For an event $\ev$ which is
  completely determined by the random variable $Y$, i.e.~for all
  $y$, the probability $\Pr[\ev | Y=y]$ either vanishes or
  equals one, we can decompose the density matrix
  $\rho_{\MC{X}{Y}{E}}$ into\footnote{One is tempted to think that such a decomposition holds for {\em any} event $\ev$; however, this is not true. See
  Lemma~2.1 of~\cite{DFSS07} for another special case where the decomposition does hold.}
\[ \rho_{\MC{X}{Y}{E}} = \Pr[\ev] \cdot \rho_{\MC{X}{Y}{E}|\ev} +
\Pr[\overline{\ev}] \cdot \rho_{\MC{X}{Y}{E}|\overline{\ev}} \, .
\]
\end{lemma}

\section{Protocols and Functionalities}\label{sec:prots}

\subsubsection{Quantum Protocols.}

We consider {\em two-party quantum protocols $\qp = (\A,\B)$},
consisting of interactive quantum algorithms $\A$ and $\B$. For
convenience, we call the two parties who run $\A$ and $\B$ {\em
  Alice} and {\em Bob}, respectively.  There are different approaches
to formally define interactive quantum algorithms and thus quantum
two-party protocols, in particular when we restrict in- and outputs
(of honest participants) to be classical. For instance such a formalization can be
done by means of quantum circuits, or by means of a classical Turing
machine which outputs unitaries that are applied to a quantum
register.  For our work, the specific choice of the formalization is
immaterial; what is important is that such a two-party quantum
protocol, formalized in whatever way, uniquely specifies its
input-output behavior. Therefore, in this work, we capture quantum
protocols by their input-output behavior, which we formalize by a
quantum operation, i.e.\ a trace-preserving completely-positive map,
which maps the common two-partite input state $\rho^\nord_{UV}$ to the
common two-partite output state $\rho_{XY}$. We denote this operation by
$\rho_{XY} = \qp \, \rho^\nord_{UV}$ or, when we want to emphasize
that $\qp$ is executed by {\em honest} Alice and Bob, also by
$\rho_{XY} = \qp_{\A,\B} \, \rho^\nord_{UV}$. If one of the players,
say Bob, is {\em dishonest} and follows a malicious strategy $\dB$,
then we slightly abuse notation and write $\qp_{\A,\dB}$ for the corresponding operator.

\vspace{-2ex}\subsubsection{Protocols and Functionalities with Classical In- and Output.}

In this work, we focus on quantum protocols $\qp = (\A,\B)$ with {\em
  classical in- and output} for the honest players. This means that
we assume the common input state $\rho^\nord_{UV}$ to be classical,
i.e.\ of the form $\rho^\nord_{UV} = \sum_{u,v} P_{UV}(u,v)
\proj{u}\otimes\proj{v}$ for some probability distribution $P_{UV}$,
and the common output state $\rho_{XY} = \qp_{\A,\B} \,
\rho^\nord_{UV}$ is then guaranteed to be classical as well, i.e.,
$\rho_{XY} = \sum_{x,y} P_{XY}(x,y) \proj{x} \otimes \proj{y}$.  In
this case we may understand $U$ and $V$ as well as $X$ and $Y$ as
random variables, and we also write $(X,Y) = \qp(U,V)$.  Note that the
input-output behavior of the protocol is uniquely determined by the
conditional probability distribution $P_{XY|UV}$.  If one of the
players, say Bob, is dishonest and follows a malicious strategy $\dB$,
then we may allow his part of the input to be quantum and denote it as
$V'$, i.e. $\rho^\nord_{UV'} = \sum_{u} P_{U}(u)
\proj{u}\otimes\rho^\nord_{V'|U=u}$, and we allow his part $Y'$ of the
common output state $\rho_{XY'} = \qp_{\A,\dB} \, \rho^\nord_{UV'}$ to
be quantum, i.e. $\rho_{XY'} = \sum_{x} P_{X}(x)
\proj{x}\otimes\rho_{Y'|X=x}$. We write $\rho^\nord_{UV'}$ as
$\rho^\nord_{U\emptyset} = \rho^\nord_{U} \otimes \rho^\nord_{\emptyset} = \rho^\nord_U$ if $V'$ is empty, i.e.\
if $\dB$ has no input at all, and we write it as $\rho^\nord_{UZV'}$
if part of his input, $Z$, is actually classical.

A classical non-reactive two-party {\em ideal functionality} $\F$ is
given by a conditional probability distribution $P_{\F(U,V)|UV}$,
inducing a pair of random variables $(X,Y) = \F(U,V)$ for every joint
distribution of $U$ and $V$.  We also want to take into account ideal
functionalities which allow the dishonest player some
additional---though still limited---possibilities (as for instance in Section~\ref{sec:ident} or~\ref{sec:OT}). We do this as follows. We specify $\F$ not only
for the ``proper'' domains $\cal U$ and $\cal V$, over which $U$ and
$V$ are supposed to be distributed, but we actually specify it for
some larger domains $\tilde{\cal U} \supseteq {\cal U}$ and
$\tilde{\cal V} \supseteq {\cal V}$. The understanding is that $U$ and
$V$ provided by honest players always lie in $\cal U$ and
$\cal V$, respectively, whereas a dishonest player, say Bob, may
select $V$ from $\tilde{\cal V} \setminus {\cal V}$, and this way Bob
may cause $\F$, if specified that way, to process its inputs
differently and/or to provide a ``more informative'' output $Y$ to
Bob.  For simplicity though, we often leave the possibly different
domains for honest and dishonest players implicit.

We write $(X,Y) = \F_{\hA,\hB}(U,V)$ or $\rho_{XY} = \F_{\hA,\hB} \,
\rho^\nord_{UV}$ for the execution of the ``ideal-life'' protocol,
where Alice and Bob forward their inputs to $\F$ and output whatever
they obtain from~$\F$.  And we write \smash{$\rho_{XY'} =
  \F_{\hA,\dhB} \, \rho^\nord_{UV'}$} for the execution of this
protocol with a dishonest Bob with strategy $\dhB$ and quantum input
$V'$. Note that Bob's possibilities are very limited: he can produce
some classical input $V$ for $\F$ (distributed over $\tilde{\cal V}$)
from his input quantum state $V'$, and then he can prepare and output
a quantum state $Y'$ which might depend on $\F$'s reply $Y$.

\vspace{-2ex}\subsubsection{Classical Hybrid Protocols.}
A two-party {\em classical hybrid} protocol $\cp^{\F_1\cdots\F_\ell} =
(\hA,\hB)$ between Alice and Bob is a protocol which makes a bounded
number $k$ of sequential oracle calls to possibly different
ideal functionalities $\F_1,\ldots,\F_\ell$.  We allow $\hA$ and $\hB$
to make several calls to independent copies of the same $\F_i$, but
we require from $\cp^{\F_1\cdots\F_\ell}$ that for every possible
execution, there is always agreement between $\hA$ and $\hB$ on when
to call which functionality; for instance we may assume that $\hA$ and
$\hB$ exchange the index $i$ before they call $\F_i$ (and stop if
there is disagreement). 

\setlength{\intextsep}{4mm}
\begin{wrapfigure}{R}{56mm}
\smallskip
\hfill\PDForPSinclude{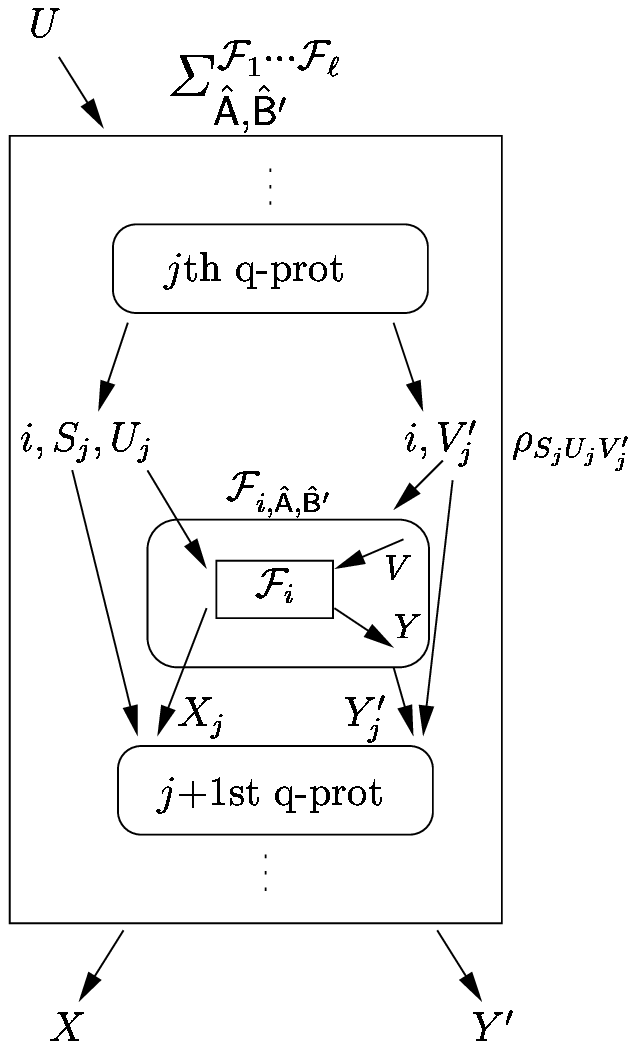}{5cm}
\begin{center}
\hfill\begin{minipage}{50mm}\small
\caption{Hybrid protocol $\cp^{\F_1\cdots\F_\ell}_{\hA,\dhB}$} \label{fig:hybridprotocol} 
\end{minipage} $ $
\vspace{-3ex}
\end{center}
\end{wrapfigure}
Formally, such a classical hybrid protocol is given by a sequence of
$k+1$ quantum protocols formalized by quantum operators with classical
in- and output for the honest players, see
Figure~\ref{fig:hybridprotocol}.  For an honest player, say Alice, the
$j$-th protocol outputs an index $i$ indicating which functionality is
to be called, classical auxiliary (or ``state'') information
information $S_j$ and a classical input $U_j$ for $\F_i$. The
$(j+1)$-st protocol expects as input $S_j$ and Alice's classical
output $X_j$ from $\F_i$. Furthermore, the first protocol expects
Alice's classical input $U$ to the hybrid protocol, and the last
produces the classical output $X$ of the hybrid protocol. In case of a
dishonest player, say Bob, all in- and outputs may be quantum states
$V'_j$ respectively $Y'_j$.  By instantiating the $j$-th call to a
functionality $\F$ (where we from now on omit the index for simpler
notation) in the obvious way by the corresponding ``ideal-life''
protocol $\F_{\hA,\hB}$ (respectively \smash{$\F_{\dhA,\hB}$} or
\smash{$\F_{\hA,\dhB}$} in case of a dishonest Alice or Bob), we
obtain the instantiated hybrid protocol formally described by quantum
operator $\cp^{\F_1\cdots\F_\ell}_{\hA,\hB}$ (respectively
$\cp^{\F_1\cdots\F_\ell}_{\dhA,\hB}$ or
$\cp^{\F_1\cdots\F_\ell}_{\hA,\dhB}$).\footnote{Note that for simpler
  notation, we are a bit sloppy and give the same name, like $\hA$ and
  $\dhB$, to honest Alice's and dishonest Bob's strategy within
  different (sub)protocols.  }

For the hybrid protocol to be {\em classical}, we mean that it has
classical in- and output (for the honest players), but also that all
communication between Alice and Bob is classical.\footnote{We do not
  explicitly require the internal computations of the honest parties
  to be classical. } Since we have not formally modeled the
communication within (hybrid) protocols, we need to formalize this
property as a property of the quantum operators that describe the
hybrid protocol: Consider a dishonest player, say Bob, with no input,
and consider the common state \smash{$\rho_{S_j U_j V'_j}$} at any
point during the execution of the hybrid protocol when a call to
functionality $\F_i$ is made. The requirement for the hybrid protocol
to be \emph{classical} is now expressed in that there exists a
classical $Z_j$---to be understood as consisting of $\dhB$'s classical
communication with $\hA$ and with the $\F_{i'}$'s up to this
point---such that given $Z_j$, Bob's quantum state $V'_j$ is uncorrelated with (i.e.~independent of) Alice' classical input and auxiliary information:
\smash{$\rho_{S_j U_j Z_j V'_j} = \rho_{\MC{S_j U_j}{Z_j}{V'_j}}$}.
Furthermore, we require that we may assume $Z_j$ to be part of $V'_j$
in the sense that for any $\dhB$ there exists $\dhB'$ such that $Z_j$
is part of $V'_j$. This definition is motivated by the observation
that if Bob can communicate only classically with Alice, then he can
correlate his quantum state with
information on Alice's side only by means of the classical communication.

We also consider the protocol we obtain by replacing the ideal
functionalities by quantum two-party sub-protocols
$\qp_1,\ldots,\qp_\ell$ with classical in- and outputs for the honest
parties: whenever $\cp^{\F_1\cdots\F_\ell}$ instructs $\hA$ and $\hB$
to execute ${\F_i}_{\hA,\hB}$, they instead execute $\qp_i =
(\A_i,\B_i)$ and take the resulting outputs. We write
$\cp^{\qp_1\cdots\qp_\ell} = (\A,\B)$ for the real quantum
protocol we obtain this way.

\section{Security for Two-Party Quantum Protocols}

\subsection{The Security Definition}\label{sec:SecDef}

\subsubsection{Framework.}
We use the following framework for defining security of a quantum
protocol~$\qp$ with classical in- and output. We distinguish three
cases and consider the respective output states obtained by executing
$\qp$ in case of honest Alice and honest Bob, in case of honest Alice
and dishonest Bob, and in case of dishonest Alice and honest Bob. For
each of these cases we require some security conditions on the output
state to hold. More precisely, for honest Alice and Bob, we fix an
arbitrary joint probability distribution $P_{UV}$ for the inputs $U$
and $V$, resulting in outputs $(X,Y) = \qp_{\A,\B}(U,V)$ with a well
defined joint probability distribution $P_{UVXY}$. For an honest Alice
and a dishonest Bob, we fix an arbitrary distribution $P_U$ for
Alice's input and an arbitrary strategy $\dB$ {\em with no input} for
Bob, and we consider the resulting joint output state
$$
\rho_{UXY'} = \big(\id_U \otimes \qp_{\A,\dB}\big) \rho_{UU\emptyset}
= \sum_u
P_U(u)\proj{u}\otimes\qp_{\A,\dB}(\proj{u}\!\otimes\!\rho_\emptyset)
$$ 
augmented with Alice's input $U$, where $U$ and $X$ are classical and
$Y'$ is in general quantum. And, correspondingly, for a dishonest
Alice and an honest Bob, we fix an arbitrary distribution $P_V$ for
Bob's input and an arbitrary strategy $\dA$ {\em with no input} for
Alice, and we consider the resulting joint output state
$$
\rho_{VX'Y} = \big(\id_V \otimes \qp_{\dA,\B}\big) \rho_{V\emptyset V}
= \sum_v
P_V(v)\proj{v}\otimes\qp_{\dA,\B}(\rho_\emptyset\!\otimes\!\proj{v})
$$ 
augmented with Bob's input $V$.  Then, security is defined by specific
information-theoretic conditions on $P_{UVXY}$, $\rho_{UXY'}$ and
$\rho_{VX'Y}$, where the conditions depend on the functionality $\F$
which $\qp$ is implementing. Definition~\ref{def:security}
below for a general functionality $\F$, as well as the definitions
studied later for specific functionalities
(Definitions~\ref{def:ident},~\ref{def:Rl12OT} etc.), are to be
understood in this framework. In particular, the augmented common
output states are to be understood as defined above.

We stress once more that the framework assumes that dishonest players
have no input at all. This might appear too weak at first glance; one
would expect a dishonest player, say Bob, to at least get the input
$V$ of the honest Bob. The justification for giving dishonest players
no input is that on the one hand, we will show that this ``minimalistic
approach'' is good enough for the level of security we are aiming for
(see~Theorem~\ref{thm:Comp}), and on the other hand, our goal is to
keep the security definitions as simple as possible.

\vspace{-2ex}\subsubsection{Restricting the Adversary.}

Since essentially no interesting two-party task can be implemented
securely by a quantum protocol against unbounded quantum
attacks \cite{Mayers97,LC97,Lo97,Kitaev03}, one typically has to put
some restriction upon the dishonest player's capabilities.  One such
restriction, which proved to lead to interesting results, is to limit
the quantum-storage capabilities of the dishonest
player~\cite{DFSS05,DFRSS07,DFSS07,WST07arxiv}, but one can also
consider other restrictions like a bound on the size of coherent
measurements dishonest players can do~\cite{Salvail98}.

Throughout, we let $\mathfrak{A}$ and $\mathfrak{B}$ be subfamilies of all
possible strategies $\dA$ and $\dB$ of a dishonest Alice and a
dishonest Bob, respectively.  In order to circumvent some pathological
counter examples, we need to assume the following two natural
consistency conditions on $\mathfrak{A}$, and correspondingly on
$\mathfrak{B}$. If a dishonest strategy $\dA \in \mathfrak{A}$ expects as
input some state $\rho_{ZU'}$ with classical $Z$, then for any $z$ and
for any $\rho_{U'|Z=z}$, the strategy $\dA_{z,\rho_{U'|Z=z}}$, which
has $z$ hard-wired and prepares the state $\rho_{U'|Z=z}$ as an
initial step but otherwise runs like $\dA$, is in $\mathfrak{A}$ as well.
And, if $\dA \in \mathfrak{A}$ is a dishonest strategy for a protocol
$\cp^\qp$ which makes a call to a sub-protocol $\qp$, then the
corresponding ``sub-strategy'' of $\dA$, which is active during the
execution of $\qp$, is in $\mathfrak{A}$ as well. 

\vspace{-2ex}\subsubsection{Defining Security.}

Following the framework described above, we propose the following security definition for two-party quantum protocols with classical in- and output. The justification for the proposed definition is that it implies strong simulation-based security when using quantum protocols as sub-protocols in classical outer protocols (Theorem~\ref{thm:Comp}), yet the definition is expressed in a way that is as simple and as weak as (seemingly) possible, making it as easy as possible to design and prove quantum cryptographic schemes secure according to the definition. 

\begin{definition}
\label{def:security}
A two-party quantum protocol $\qp$ {\em $\eps$-securely implements} an ideal classical functionality $\F$ against $\mathfrak{A}$ and $\mathfrak{B}$ if the following holds:
\begin{description}
\item[{\em Correctness:}] For any joint distribution of the input $U$ and $V$, the resulting common output $(X,Y) = \qp(U,V)$ satisfies 
$$(U,V,X,Y) \epsclose (U,V,\F(U,V)) \, .$$
\item[{\em Security for Alice:}] For any $\dB \in \mathfrak{B}$ (with
  no input), and for any distribution of $U$, the resulting common
  output state $\rho_{UXY'}$ (augmented with $U$) is such that there
  exist\footnote{as defined in Section~\ref{sec:notation}.}
  classical random variables $V$ and $Y$ such that
$$
P_{UV} \epsclose P_U \cdot P_V, \;\; 
(U,V,X,Y) \epsclose (U,V,\F(U,V)) \;\;\text{and}\;\;
\rho_{UXVYY'} \epsclose \rho_{\MC{UX}{VY}{Y'}} \, .
$$
\item[{\em Security for Bob:}] For any $\dA \in \mathfrak{A}$ (with no
  input), and for any distribution of $V$, the resulting common output
  state $\rho_{VX'Y}$ (augmented with $V$) is such that there
  exist
  classical random variables $U$ and $X$ such that
$$P_{UV} \epsclose P_U \cdot P_V, \;\;
(U,V,X,Y) \epsclose (U,V,\F(U,V))\;\;\text{and}\;\;
\rho_{VYUXX'} \epsclose \rho_{\MC{VY}{UX}{X'}} \, .
$$
\end{description}
\end{definition}
The three conditions for dishonest Bob (and similarly for dishonest
Alice) express that, up to a small error, $V$ is independent of $U$,
$X$ and $Y$ are obtained by applying $\F$, and the quantum state $Y'$
is obtained by locally processing $V$ and $Y$.

\subsection{Equivalent Formulations}\label{sec:discussion}

As already mentioned, Definition~\ref{def:security} appears to
guarantee security only in a very restricted setting, where the honest
player has no information beyond his input, and the dishonest player
has no (auxiliary) information at all. Below, we argue that
Definition~\ref{def:security} actually implies security in a somewhat
more general setting, where the dishonest player is allowed as input
to have arbitrary classical information $Z$ as well as a quantum state
which only depends on $Z$.  For completeness, although this is rather
clear, we also argue that not only the honest player's input is
protected, but also any classical ``side information'' $S$ he might
additionally have but does not use. 

\begin{proposition} \label{prop:sideinfo}
Let $\qp$ be a two-party protocol that $\eps$-securely implements $\F$ against $\mathfrak{A}$ and $\mathfrak{B}$. 
Let $\dB \in \mathfrak{B}$ be a dishonest Bob who takes as input a classical $Z$ and a quantum state $V'$ 
and outputs (the same) $Z$ and a quantum state $Y'$. Then, 
for any $\rho^\nord_{SUZV'}$ with $\rho^\nord_{SUZV'} = \rho^\nord_{\MC{SU}{Z}{V'}}$, the resulting overall output state (augmented with $S$ and $U$) 
$$
\rho_{SUXZY'} = \big(\id_{SU} \otimes \qp_{\A,\dB}\big) \rho^\nord_{SUUZV'} = \sum_{s,u,z} P_{SUZ}(s,u,z)\proj{s,u}\otimes\qp_{\A,\dB}(\proj{u}\otimes\proj{z}\otimes\rho^\nord_{V'|Z=z})
$$ 
is such that there exist classical random variables $V$ and $Y$ such that  
$P_{SUZV} \epsclose P_{\MC{SU}{Z}{V}}$, 
$(S,U,V,X,Y,Z) \epsclose (S,U,V,\F(U,V),Z)$
and
$\rho_{SUXVYZY'} = \rho_{\MC{SUX}{VYZ}{Y'}}$. 
The corresponding holds for a dishonest Alice. 
\end{proposition}
\begin{proof}
It is rather clear that we can extend the setting from
Definition~\ref{def:security} by $S$: We can view $S$ as an
additional input to $\F$, provided by Alice besides $U$, which is
simply ignored by $\F$. Definition~\ref{def:security} then
immediately implies that the common output state $\rho_{SUXY'}$
allows $V$ and $Y$ such that $P_{SUV} \epsclose P_{SU} P_V$,
$(S,U,V,X,Y) \epsclose (S,U,V,\F(U,V))$ and $\rho_{SUXVYY'} \epsclose
\rho_{\MC{SUX}{VY}{Y'}}$.
  
Consider now a dishonest Bob who holds some classical auxiliary
information $Z$. Applying Definition~\ref{def:security} with the
above observation to the distribution $P_{SU|Z=z}$ and the dishonest
Bob who has $z$ hard-wired and locally prepares
$\rho^\nord_{V'|S=s,U=u,Z=z} = \rho^\nord_{V'|Z=z}$ implies
that the conditioned common output state $\rho_{SUXY'|Z=z}$ allows $V$
and $Y$ such that $P_{SUV|Z=z} \epsclose P_{SU|Z=z} P_{V|Z=z}$,
$P_{SUVXY|Z=z} \epsclose P_{SUV\F(U,V)|Z=z}$ and $\rho_{SUXVYV'|Z=z}
\epsclose \rho_{\MC{SUX}{VY}{V'|Z=z}}$.
As the above holds for any $z$, it follows that $P_{SUZV} \epsclose P_{\MC{SU}{Z}{V}}$, $P_{SUVXYZ} \epsclose P_{SUV\F(U,V)Z}$ as well as that
\begin{align*}
\rho_{SUXVYZY'} &= \sum_z P_Z(z) \proj{z} \otimes \rho_{SUXVYY'|Z=z} \\
&\epsclose \sum_z P_Z(z) \proj{z} \otimes  \rho_{\MC{SUX}{VY}{Y'|Z=z}}  \\
&= \sum_z P_Z(z) \proj{z} \otimes \sum_{suxvy} P_{SUXVY|Z}(s,u,x,v,y|z)\proj{suxvy} \rho_{Y'}^{vyz} \\ 
&= \sum_{suxvyz} P_{SUXVYZ}(s,u,x,v,y,z)\proj{suxvyz} \rho_{Y'}^{vyz} \\
&= \rho_{\MC{SUX}{VYZ}{Y'}} \, .
\end{align*} 

\vspace{-3ex}
\qed 
\end{proof}

Note the restriction on the adversary's quantum input $V'$, namely that it
is only allowed to depend on the honest player's input $U$ (and side
information $S$) ``through'' $Z$. It is this limitation which
prohibits quantum protocols satisfying Definition~\ref{def:security}
to securely compose into outer quantum protocols but requires the
outer protocol to be classical. Indeed, within a quantum protocol that
uses quantum communication, a dishonest player may be able to correlate
his quantum state with classical information on the honest player's
side; however, within a classical protocol, he can only do so through
the classical communication so that his state is still independent
when given the classical communication.

The following proposition 
shows equivalence to a simulation-based definition; this will be a
handy formulation in order to prove the composition theorem.

\begin{proposition}\label{prop:simulate}
Let $\qp$ be a two-party protocol that $\eps$-securely implements $\F$ against $\mathfrak{A}$ and $\mathfrak{B}$. 
Let $\dB \in \mathfrak{B}$ be a dishonest Bob who takes as input a classical $Z$ and a quantum state $V'$, engages into $\qp$ with honest Alice and outputs $Z$ and a quantum state $Y'$. Then, 
for any $\rho^\nord_{SUZV'}$ with $\rho^\nord_{SUZV'} = \rho^\nord_{\MC{SU}{Z}{V'}}$ there exists \smash{$\dhB$} such that
$$
\big(\id_{S} \otimes \qp_{\A,\dB}\big) \rho^\nord_{SUZV'}
\epsclose[3\eps] \big(\id_{S} \otimes \F_{\hA,\dhB}\big)
\rho^\nord_{SUZV'} \, .$$
The corresponding holds for a dishonest Alice. 
\end{proposition}
\begin{proof}
Given that $Z = z$, $\dhB$ samples $v$ according to the distribution $P_{V|Z=z}$, and sends it to $\F$ in order to receive output $y$. Then, $\dhB$ prepares and outputs the quantum state $\rho_{Y'}^{vyz}$. The resulting common output state $\hat{\rho}_{SUXZY'}$ (augmented with $S$ and $U$) is as follows. 
\begin{align*}
\hat{\rho}_{SUXZY'} &= \sum_{s,u,z} P_{SUZ}(s,u,z) \sum_v P_{V|Z}(v|z) \sum_{x,y} P_{\F(U,V)|UV}(x,y|u,v)  \proj{s,u,x,z} \rho_{Y'}^{vyz} \\
&\epsclose \sum_{s,u,v,z} P_{SUVZ}(s,u,v,z) \sum_{x,y} P_{\F(U,V)|UV}(x,y|u,v) \proj{s,u,x,z} \rho_{Y'}^{vyz} \\
&= \sum_{s,u,v,z} P_{SUVZ}(s,u,v,z) \sum_{x,y} P_{\F(U,V)|SUVZ}(x,y|s,u,v,z)  \proj{s,u,x,z} \rho_{Y'}^{vyz} \\
&\epsclose \sum_{s,u,v,x,y,z} P_{SUVXYZ}(s,u,v,x,y,z) \proj{s,u,x,z} \rho_{Y'}^{vyz} \\
&= \rho_{\MC{SUX}{Z}{Y'}} 
\epsclose \rho_{SUXZY'} \, .
\end{align*}

\vspace{-3ex}
\qed
\end{proof}
Recall that \smash{$\F_{\hA,\dhB}$} is the execution of the ``ideal-life'' protocol, where honest $\hA$ relays in- and outputs, and the only thing dishonest $\dhB$ can do is modify the input and the output. 
Note that we do not guarantee that $\dhB$ is in $\mathfrak{B}$; we will comment on this after Theorem~\ref{thm:Comp}.

\section{Composability}

We show the following composition result. If quantum protocols
$\qp_1,\ldots,\qp_\ell$ securely implement ideal functionalities
$\F_1,\ldots,\F_\ell$ according to Definition~\ref{def:security}, then any
two-party {\em classical} hybrid protocol $\cp^{\F_1,\ldots,\F_\ell}$
which makes sequential calls to $\F_1,\ldots,\F_\ell$ is essentially
equally secure as the protocol obtained by replacing the calls to
$\F_1,\ldots,\F_\ell$ by the respective quantum subroutines
$\qp_1,\ldots,\qp_\ell$.

We stress that the $\F_i$'s are {\em classical} functionalities, i.e.,
even a dishonest player $\dhA$ or $\dhB$ can only input a classical value
to $\F_i$, and for instance cannot execute $\F_i$ with several
inputs in superposition.
This makes our composition result stronger, because we give the adversary less power in the ``ideal'' (actually hybrid) world.

\begin{theorem}[Composition Theorem]\label{thm:Comp}
 Let $\cp^{\F_1\cdots\F_\ell} = (\hA,\hB)$ be a classical two-party hybrid protocol
 which makes at most $k$ oracle calls to the functionalities, and for every $i \in \set{1,\ldots,\ell}$, let protocol $\qp_i$ be an $\eps$-secure implementation of $\F_i$ against $\mathfrak{A}$ and $\mathfrak{B}$. Then, the following holds. 
 \begin{description}
\item[{\em Correctness:}]  For every distribution of $U$ and  $V$
$$
\delta\Bigl( \cp^{\qp_1\cdots\qp_\ell}_{\A,\B} \rho^\nord_{U V},
\cp^{\F_1\cdots\F_\ell}_{\hA,\hB} \rho^\nord_{U V} \Bigr) \leq k
\eps \, .
$$
\item[{\em Security for Alice:}] For every $\dB \in \mathfrak{B}$ there
exists $\dhB$ such that for every distribution of~$U$ 
$$
\delta\Bigl( \cp^{\qp_1\cdots\qp_\ell}_{\A,\dB} \rho^\nord_{U\emptyset},
\cp^{\F_1\cdots\F_\ell}_{\hA,\dhB} \rho^\nord_{U\emptyset} \Bigr) \leq 3 k
\eps \, .
$$
\item[{\em Security for Bob:}] For every $\dA \in \mathfrak{A}$ there exists $\dhA$ such that for every distribution of~$V$  
$$
\delta\Bigl( \cp^{\qp_1\cdots\qp_\ell}_{\dA,\B} \rho^\nord_{\emptyset V},
\cp^{\F_1\cdots\F_\ell}_{\dhA,\hB} \rho^\nord_{\emptyset V} \Bigr) \leq 3 k
\eps \, .
$$
\end{description}
\end{theorem}
Before going into the proof, we would like to point out the following observations. 
First of all, note that the quantification is such that the dishonest hybrid adversary
$\dhB$ (and correspondingly $\dhA$) does {\em not} depend on the distribution of the honest player's input $U$, and as such we do not need to assume that the adversary knows the honest player's input distribution.

Also note that in contrast to typical composition theorems, which
per-se guarantee security when replacing {\em one} functionality by a
sub-protocol and where in case of several functionalities security
then follows by induction, Theorem~\ref{thm:Comp} is stated in such a
way that it directly guarantees security when replacing all
functionalities by sub-protocols. The reason for this is that the
assumption that the outer protocol is classical is not satisfied
anymore once the first functionality is replaced by a quantum
sub-protocol, and thus the inductive reasoning does not work directly.
We stress that our composition theorem nevertheless allows for several
levels of compositions (see Corollary~\ref{cor:comp} and the
preceding discussion).

Furthermore, note that we do not guarantee that the dishonest hybrid adversary
$\dhB$ is in $\mathfrak{B}$ (and similarly for~$\dhA$). For instance the
specific $\dhB$ we construct in the proof is more involved with
respect to classical resources (memory and computation), but less
involved with respect to quantum resources: essentially it follows $\dB$, except
that it remembers all classical communication and except that the actions during
the sub-protocols are replaced by sampling a value from some
distribution and preparing a quantum state (of a size that also $\dB$
has to handle); the descriptions of the distribution and the state
have to be computed by $\dhB$ from the stored classical communication.
By this, natural restrictions on $\dB$ concerning its {\em quantum
  capabilities} propagate to $\dhB$. For instance if $\dB$ has a
quantum memory of bounded size, so has $\dhB$.  Furthermore, in many
cases the classical hybrid protocol is actually {\em unconditionally}
secure against classical dishonest players and as such in particular
secure against unbounded quantum dishonest players (because every dishonest quantum
strategy can be simulated by an unbounded classical
adversary), so no restriction on $\dhB$ is needed.

Finally, note that we do not specify what it means for the hybrid
protocol to be secure; Theorem~\ref{thm:Comp} guarantees that {\em
  whatever} the hybrid protocol achieves, essentially the same is achieved by the
real-life protocol with the oracle calls replaced by protocols.  But
of course in particular, if the hybrid protocol {\em is} secure in the
sense of Definition~\ref{def:security}, then so is the real-life
protocol, and as such it could itself be used as a quantum sub-protocol in yet another classical outer protocol. 

\begin{corollary}\label{cor:comp}
If $\cp^{\F_1\cdots\F_\ell}$ is a $\delta$-secure implementation of $\cal G$ against $\mathfrak{A}$ and $\mathfrak{B}$, and if $\qp_i$ is an $\eps$-secure implementation of $\F_i$ against $\mathfrak{A}$ and $\mathfrak{B}$ for every $i \in \set{1,\ldots,\ell}$, then $\cp^{\qp_1\cdots\qp_\ell}$ is a $(\delta\!+\!3k\eps)$-secure implementation of~$\cal G$. 
\end{corollary}

\def\mark#1{\bar{#1}}

\begin{proof}[of Theorem~\ref{thm:Comp}]
Correctness is obvious. We show security for Alice; security for Bob can be shown accordingly. 
Consider a dishonest $\dB$. 
First we argue that for every distribution for Alice's input $U$, there exists a $\dhB$ as claimed (which though may depend on $P_U$). Then, in the end, we show how to make $\dhB$ independent of $P_U$. 

Let $\A$'s input $U$ be
arbitrarily distributed. We prove the claim by induction on
$k$. The claim holds trivially for protocols that make zero oracle
calls.  Consider now a protocol $\cp^{\F_1\cdots\F_\ell}$ with at
most $k > 0$ oracle calls.  For simplicity, we assume that the
number of oracle calls equals $k$, otherwise we instruct the players
to makes some ``dummy calls''.  Let $\rho^\nord_{S_k U_k V'_k}$ be
the common state right before the $k$-th and thus last call to one
of the sub-protocols $\qp_1,\ldots,\qp_\ell$ in the execution of the
real protocol $\cp^{\qp_1,\ldots,\qp_\ell}$. To simplify notation in
the rest of the proof, we omit the index $k$ and write \smash{$\rho^\nord_{\mark{S}
\mark{U} \mark{V}'}$} instead; see Figure~\ref{fig:compproof}. We
know from the induction hypothesis for $k-1$ that there exists
$\dhB$ such that $\rho^\nord_{\mark{S} \mark{U} \mark{V}'}
\approx_{3(k-1)\eps} \sigma^\nord_{\mark{S} \mark{U} \mark{V}'}$
where $\sigma^\nord_{\mark{S} \mark{U} \mark{V}'}$ is the common
state right before the $k$-th call to a functionality in the
execution of the hybrid protocol
\smash{$\cp^{\F_1\cdots\F_\ell}_{\hA,\dhB} \rho_{U\emptyset}$}. As
described in Section~\ref{sec:prots}, $\mark{S}, \mark{U}$ and
$\mark{V}'$ are to be understood as follows.  $\mark{S}$ denotes
$\A$'s (respectively \smash{$\hA$'s}) classical auxiliary
information to be ``remembered'' during the call to the
functionality. $\mark{U}$ denotes $\A$'s (respectively $\hA$'s)
input to the sub-protocol (respectively functionality) that is to be
called next, and $\mark{V}'$ denotes the dishonest player's current
quantum state.  For simplicity, we assume that the index $i$, which
determines the sub-protocol $\qp_i$ (functionality~$\F_i$) to be
called next, is {\em fixed} 
and we just write $\qp$ and $\F$ for $\qp_i$ and $\F_i$,
respectively. If this is not the case, we consider
$\rho^\nord_{\mark{S}\mark{U}\mark{V}'|\mark{I}=i}$ and
$\sigma^\nord_{\mark{S}\mark{U}\mark{V}'|\mark{I}=i}$ instead, and
reason as below for any $i$, where $\mark{I}$ denotes the index of
the sub-protocol (functionality) to be called. Note that
conditioning on $\mark{I} = i$ means that we allow $\dhB$ to depend
on $i$, but this is legitimate since $\mark{I}$ is known to the
dishonest party.

\begin{figure}
 \centering 
\PDForPSinclude{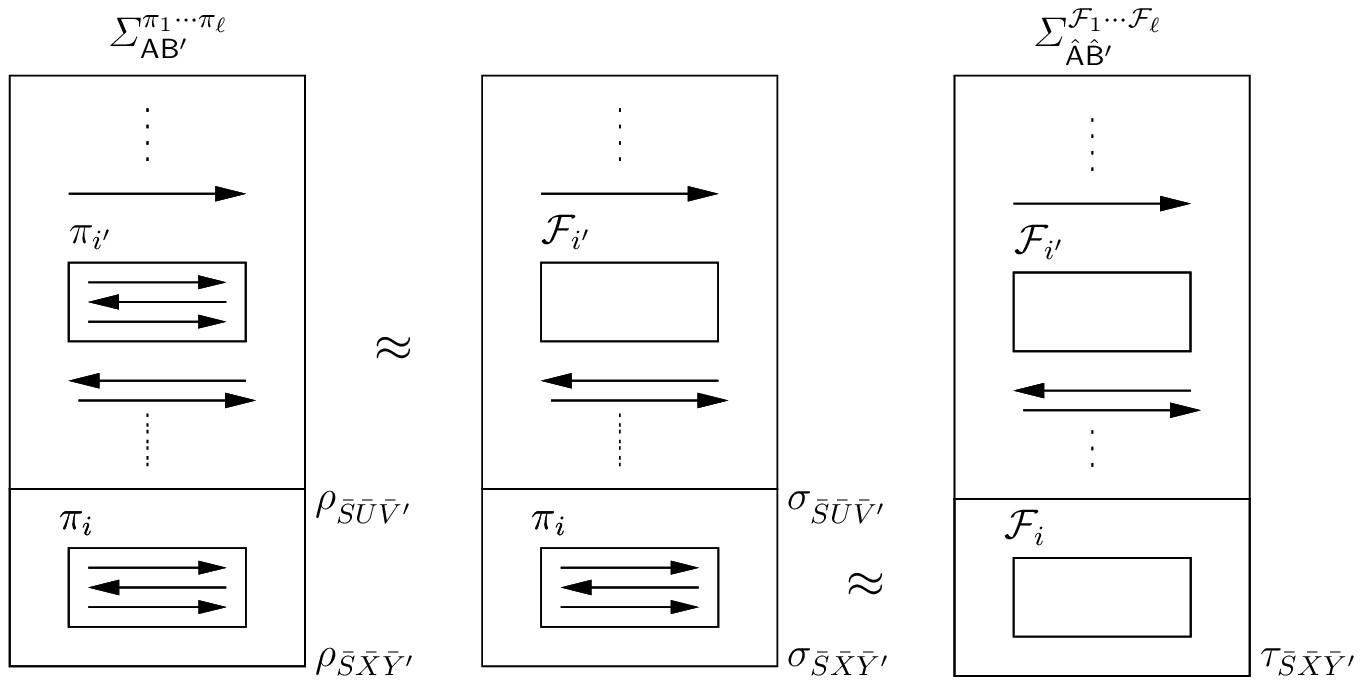}{10.5cm}
\small
 \caption{Steps of the Composability Proof} 
 \label{fig:compproof} 
\end{figure}

Consider now the evolution of the state \smash{$\sigma^\nord_{\mark{S}
  \mark{U} \mark{V}'}$} when executing \smash{$\F_{\hA,\dhB}$} (as
prescribed by the hybrid protocol) with a strategy for $\dhB$ yet to
be determined and when executing $\qp_{\A,\dB}$ instead. Let
$\sigma_{\mark{S}\mark{X}\mark{Y}'}$ and
$\tau_{\mark{S}\mark{X}\mark{Y}'}$ denote the corresponding states
after the execution of respectively $\qp_{\A,\dB}$ and
$\F_{\hA,\dhB}$, see Figure~\ref{fig:compproof}. We show that $\sigma_{\mark{S}\mark{X}\mark{Y}'}$ and
$\tau_{\mark{S}\mark{X}\mark{Y}'}$ are $3 \eps$-close; this then
proves the result by the fact that evolution does not increase the
trace distance and by the triangle inequality:
\begin{align*}
\rho_{\mark{S}\mark{X}\mark{Y}'} = (\id_{\mark{S}} \otimes\qp_{\A,\dB})\,
\rho^\nord_{\mark{S}\mark{U}\mark{V}'} &\approx_{3(k-1)\eps}
(\id_{\mark{S}}
\otimes\qp_{\A,\dB})\,\sigma^\nord_{\mark{S}\mark{U}\mark{V}'}
=\sigma_{\mark{S}\mark{X}\mark{Y}'}\\
&\approx_{3 \eps}
\tau_{\mark{S}\mark{X}\mark{Y}'} = (\id_{\mark{S}} \otimes\F_{\hA,\dhB})\,
\sigma^\nord_{\mark{S}\mark{U}\mark{V}'} \, .
\end{align*}

Let $\sigma^\nord_{\mark{S}\mark{U}\mark{Z}\mark{V}'}$,
$\sigma_{\mark{S}\mark{X}\mark{Z}\mark{Y}'}$ and
$\tau_{\mark{S}\mark{X}\mark{Z}\mark{Y}'}$ be the extensions of the
respective states $\sigma^\nord_{\mark{S}\mark{U}\mark{V}'}$,
$\sigma_{\mark{S}\mark{X}\mark{Y}'}$ and
$\tau_{\mark{S}\mark{X}\mark{Y}'}$ when we also consider $\mark{Z}$
(which collects the classical communication dictated by
$\cp^{\F_1\ldots,\F_\ell}$ as well as $\dhB$'s classical inputs to and
outputs from the previous oracle calls), which is guaranteed to exist
by our formalization of a {\em classical} hybrid protocol,
so that $\mark{Z}$ is without loss of generality contained in
$\mark{V}'$ and $\sigma^\nord_{\mark{S}\mark{U}\mark{Z}\mark{V}'} =
\sigma^\nord_{\MC{\mark{S}\mark{U}}{\mark{Z}}{\mark{V}'}}$. It thus
follows from Proposition~\ref{prop:simulate} that
$\sigma_{\mark{S}\mark{X}\mark{Z}\mark{Y}'}$ and
$\tau_{\mark{S}\mark{X}\mark{Z}\mark{Y}'}$ are $3\eps$-close for a
proper strategy of $\dhB$.  Note that the strategy of $\dhB$ may
depend on the state
$\sigma^\nord_{\mark{S}\mark{U}\mark{Z}\mark{V}'}$, but since $P_U$ as
well as \smash{$\hA$}'s behavior are fixed,
$\sigma^\nord_{\mark{S}\mark{U}\mark{Z}\mark{V}'}$ is also fixed.

It remains to argue that we can make $\dhB$ independent of $P_U$. We
use an elegant argument due to Cr\'epeau and
Wullschleger~\cite{CW08}. We know that for any $P_U$ there exists a
$\dhB$ (though depending on $P_U$) as required.  For any value $u$
that $U$ may take on, let then
$$
\eps_u = \delta\Bigl( \cp^{\qp_1\cdots\qp_\ell}_{\A,\dB} \rho^\nord_{U\emptyset|U=u},
\cp^{\F_1\cdots\F_\ell}_{\hA,\dhB} \rho^\nord_{U\emptyset|U=u} \Bigr) \, .
$$
Then, $\sum_u P_U(u) \eps_u = 3k\eps$. The $\eps_u$'s depend
on $P_U$, and thus we also write $\eps_u(P_U)$. Consider now the
function $F$ which maps an arbitrary distribution $P_U$ for $U$ to a
new distribution defined as $F(P_U)(u) \assign
\frac{1+\eps_u(P_U)}{1+3k\eps} P_U(u)$. Function $F$ is 
continuous and maps a non-empty, compact, convex set onto itself. Thus, by
Brouwer's Fixed Point Theorem, it must have a fixed point: a distribution
$P_U$ with $F(P_U) = P_U$, and thus $\eps_u(P_U) = 3k\eps$ for any
$u$. It follows that $\dhB$ which works for that particular
distribution $P_U$ in fact works for any specific value for $U$ and
so for any distribution of~$U$.
\qed
\end{proof}

\section{Example: Secure Identification}\label{sec:ident}

We show that the information-theoretic security definition proposed by Damg{\aa}rd \etal\
for their secure-identification quantum protocol in the
bounded-quantum-storage model~\cite{DFSS07} implies security in our
sense for a proper functionality $\F_{\ID}$; this guarantees composability as in Theorem~\ref{thm:Comp} for
their protocol. In Section~\ref{sec:OT} and in Appendix~\ref{app:otherOT}, we show the corresponding for the 1-2 OT
scheme~\cite{DFRSS07} and for other variants of OT.

A secure identification scheme allows a user Alice to identify herself
to server Bob by securely checking whether the supplied password
agrees with the one stored by Bob.  Specifically, on respective input
strings $W_A, W_B \in {\cal W}$ provided by Alice and Bob, the
functionality outputs the bit $Y = (W_A \eqq W_B)$ to Bob.  A
dishonest server $\dB$ should learn essentially no information on
$W_A$ beyond that he can come up with a guess $W'$ for $W_A$ and
learns whether $W' = W_A$ or not, and similarly a dishonest user $\dA$
succeeds in convincing Bob essentially only if she guesses $W_B$
correctly. If her guess is incorrect then the only thing she might
learn is that her guess is incorrect. The corresponding ideal
functionality is depicted in Figure~\ref{fig:identfunc}. Note that if
dishonest $\dA$ provides the ``correct'' input $W_A = W_B$, then $\F_{\ID}$ allows $\dA$ to learn 
this while she may still enforce Bob to reject (by setting the
``override bit'' $D$ to~$0$). In Appendix~\ref{sec:identsimpler}, we study a slightly stronger variant, which does not allow this somewhat unfair option for $\dA$.\footnote{The reason we study here the weaker version is that this corresponds to the security guaranteed by the definition proposed in~\cite{DFSS07}, as we show. }

\begin{figure}
\normalsize
\begin{framed}
{\bf Functionality} $\F_{\ID}$: 
Upon receiving strings $W_A$ and $W_B$ from user Alice and from server
Bob, $\F_\ID$ outputs the bit $W_A \eqq W_B$ to Bob. 

\smallskip
If Alice is dishonest, then she may input an additional ``override bit'' $D$. In this case, $\F_\ID$ outputs the bit $W_A \eqq W_B$ to Alice and the bit $(W_A \eqq W_B) \wedge D$ to Bob. 
\vspace{-1ex}
\end{framed}
\vspace{-2ex}
\small
\caption{The Ideal Password-Based Identification Functionality.}\label{fig:identfunc}
\vspace{-1ex}
\end{figure}

We recall the security definition from~\cite{DFSS07} for a secure
identification scheme.  The definition is in the framework described
in Section~\ref{sec:SecDef}; thus, it considers a single execution of
the protocol with an arbitrary distribution for the honest players
inputs and with no input for dishonest players, and security is
defined by information-theoretic conditions on the resulting output
states. For consistency with the above notation (and the notation used
in~\cite{DFSS07}), Alice and Bob's inputs are denoted by $W_A$ and
$W_B$, respectively, rather than $U$ and $V$. Furthermore, note that
honest Alice's output $X$ is empty: $X = \emptyset$.

\begin{definition}[Secure Identification] \label{def:ident}
A password-based quantum identification scheme is {\em $\eps$-secure} (against $\mathfrak{A}$ and $\mathfrak{B})$ if the following properties hold. 
\begin{description}
\item[{\em Correctness:}] For honest user Alice and honest server Bob,
  and for any joint input distribution $P_{W_A W_B}$, Bob learns
  whether their input is equal, except with probability~$\eps$.
\item[{\em Security for Alice:}] For any dishonest server $\dB \in
  \mathfrak{B}$, and for any distribution of $W_A$, the resulting
  common output state $\rho_{W_A Y'}$ (augmented with $W_A$) is such
  that there exists a classical $W'$ that is independent of $W_A$ and
  such that
$$
\rho_{W_A W' Y'|W_A \neq W'} \close{\eps} \rho_{W_A\leftrightarrow
  W'\leftrightarrow  Y'|W_A \neq W'}
\enspace , 
$$
\item[{\em Security for Bob:}] 
For any dishonest user $\dA \in \mathfrak{A}$, and for any distribution of $W_B$, the resulting common output state $\rho_{W_B Y X'}$ (augmented with $W_B$)
is such that there exists a classical $W'$ independent of $W_B$, such that if $W_B \neq W'$ then $Y = 1$ with probability at most $\eps$, and
$$
\rho_{W_B W' X'|W' \neq W_B} \close{\eps} \rho_{W_B\leftrightarrow W'\leftrightarrow X'| W'\neq W_B} \enspace .
$$ 
\end{description}
\end{definition}
A somewhat more natural functionality (without ``override bit'') can
be achieved by slightly strengthening the requirements of
Definition~\ref{def:ident}, see Appendix~\ref{sec:identsimpler}.

\begin{proposition} \label{prop:ident}
A quantum protocol satisfying Definition~\ref{def:ident}
$3 \eps$-securely implements the functionality $\F_{\ID}$ from
Figure~\ref{fig:identfunc} according to
Definition~\ref{def:security}. 
\end{proposition}
\begin{proof}
Correctness follows immediately.

{\em Security for Alice:} 
Consider $W'$ which is guaranteed to exist by Definition~\ref{def:ident}. 
Let us define $V = W'$ and let $Y$ be the bit $W_A \eqq W'$. By the requirement of Definition~\ref{def:ident}, $W'$ is independent of Alice's input $W_A$. Furthermore, we have that
$$
\big(W_A, W', \emp,Y\big) = \big(W_A, W', \F_\text{\ID}(W_A,W') \big)
$$
by the definition of $\F_\text{\ID}$. Finally, we note that $Y$
completely determines the event $\ev \assign \set{W_A \neq W'}$ and
therefore, we conclude using Lemma~\ref{lemma:event} that
\begin{align*}
\rho_{W_A \emp W' Y Y'} &= 
\Pr[W_A \neq W'] \cdot \rho_{W_A \emp W' Y Y'|W_A
\neq W'} + \Pr[W_A = W'] \cdot \rho_{W_A \emp W' Y Y'|W_A = W'} \\
&= \Pr[W_A \neq W'] \cdot \rho_{W_A \emp W' Y Y'|W_A
\neq W'} + \Pr[W_A = W'] \cdot \rho_{\MC{W_A}{W' Y}{Y'}|W_A = W'} \\
&\approx_\eps \Pr[W_A \neq W'] \cdot \rho_{\MC{W_A}{W' Y}{Y'}|W_A
\neq W'} + \Pr[W_A = W'] \cdot \rho_{\MC{W_A}{W' Y}{Y'}|W_A = W'} \\
&= \rho_{\MC{W_A}{W' Y}{Y'}}.
\end{align*}

{\em Security for Bob:} Consider $W'$ which is guaranteed to exist by
Definition~\ref{def:ident}.  Let us define $U$ and $X$ as follows. We
let $U = (W',D)$ where we define $D = Y$ if $W_B = W'$, and else we
choose $D$ ``freshly'' to be $0$ with probability $\Pr[Y=0|W_B = W']$
and to be $1$ otherwise.  Furthermore, we let $X = (W ' \eqq W_B)$.
Recall that by the requirement of Definition~\ref{def:ident}, $W'$ is
independent of Bob's input $W_B$. Furthermore by construction, $D = 0$
with probability $\Pr[Y=0|W_B = W']$, independent of the value of
$W_B$ (and independent of whether $W_B = W'$ or not). Thus, $U$ is
perfectly independent of $W_B$.

Since by Definition~\ref{def:ident} the probability for Bob to decide
that the inputs are equal, $Y=1$, does not exceed $\eps$ if $W_B \neq
W'$, we have that
\begin{align*}
P_{U W_B X Y} &= \Pr[W_B = W'] \cdot P_{U W_B X Y|W_B = W'} + \Pr[W_B
\neq W'] \cdot P_{U W_B X Y|W_B \neq W'}\\
&= \Pr[W_B = W'] \cdot P_{U W_B \F_\text{\ID}(U,W_B)|W_B = W'} +
\Pr[W_B \neq W'] \cdot P_{U W_B X Y|W_B \neq W'}\\
&\approx_\eps \Pr[W_B = W'] \cdot P_{U W_B \F_\text{\ID}(U,W_B)|W_B =
  W'} + \Pr[W_B \neq W'] \cdot P_{U W_B \F_\text{\ID}(U,W_B)|W_B \neq W'}\\
&= P_{U W_B \F_\text{\ID}(U,W_B)}
\end{align*}

Finally, we have 
\begin{align*}
\rho_{W_B Y U X X'} &= 
\Pr[W_B \neq W'] \cdot \rho_{W_B Y W' D X X'|W_B
\neq W'} + \Pr[W_B = W'] \cdot \rho_{W_B Y W' D X X'|W_B = W'}
\end{align*}
In the case $W_B=W'$, we have by construction that $D=Y$ and
therefore, we obtain that $\rho_{W_B Y W' D X X'|W_B = W'} =
\rho_{\MC{W_B Y}{W' D}{X X'}|W_B = W'}$. If $W_B \neq W'$, it follows
from Definition~\ref{def:ident} and the fact that $D$ is sampled
independently that $\rho_{W_B W' D X'|W' \neq W_B} \close{\eps}
\rho_{W_B\leftrightarrow W' D\leftrightarrow X'| W'\neq
  W_B}$. Furthermore, the bit $X$ is fixed to $0$ in case $W_B \neq
W'$ and we only make an error of at most $\eps$ assuming that Bob's
output $Y$ is always $0$ and therefore,
\begin{align*}
\rho_{W_B Y W' D X X'|W_B \neq W'} &\approx_\eps \rho_{W_B (Y=0) W'
  D (X=0) X'|W_B \neq W'} \\
&\approx_\eps \rho_{\MC{W_B (Y=0)}{W' D (X=0)}{X'}|W_B \neq W'}\\
&\approx_\eps \rho_{\MC{W_B Y}{W' D X}{X'}|W_B \neq W'}
\end{align*}
Putting things together, we obtain
\begin{align*}
\rho&_{W_B Y U X X'}\\ &\epsclose[3\eps]
\Pr[W_B \neq W'] \cdot \rho_{\MC{W_B Y}{W' D X}{X'}|W_B \neq W'} 
+ \Pr[W_B = W'] \cdot \rho_{\MC{W_B Y}{W' D}{X X'}|W_B = W'} \\
&= \rho_{\MC{W_B Y}{(W' D) X}{X'}} \, ,
\end{align*}
where we used Lemma~\ref{lemma:right->mid} and~\ref{lemma:event} in the last step.

\vspace{-1ex}
\qed
\end{proof}

\section{Another Example: Randomized 1-2 Oblivious Transfer}\label{sec:OT}

Figure~\ref{fig:OT-func} below shows the ideal functionality for
sender-randomized 1-2 OT. It takes no input from Alice and an input
bit $C$ from Bob, and it outputs two random $\ell$-bit
strings $S_0$ and $S_1$ to Alice and an $\ell$-bit string $Y$ which
stands for the string of his choice $S_C$ to Bob. Note that it allows
a dishonest Alice to influence the distribution of $S_0$ and
$S_1$, and a dishonest Bob to influence the distribution of $S_C$; but this is good
enough for many applications, in particular to build a regular
(non-randomized) 1-2 OT in the standard manner.

\begin{figure}
\normalsize
\begin{framed}
{\bf Functionality} $\F_\text{12ROT}$ 

\medskip{\em Honestly behaving Alice and Bob}: 
Upon receiving no input from Alice and a choice bit $C \in
\set{0,1}$ from Bob, $\F_\text{12ROT}$ samples two random
  and independent strings $S_0,S_1 \in \set{0,1}^\ell$, and sends $S_0$ and $S_1$ to Alice and $S_C$ to Bob.

\medskip{\em Honest Alice and dishonest Bob}: 
Upon receiving no input from Alice and a bit $C \in \set{0,1}$ and a string $S_C \in \set{0,1}^\ell$ from Bob, $\F_\text{12ROT}$ samples a random independent string $S_{1-C} \in \set{0,1}^\ell$, and sends $S_0$ and $S_1$ to Alice. 

\medskip{\em Dishonest Alice and honest Bob}: 
Upon receiving two strings $S_0,S_1 \in \set{0,1}^\ell$ from Alice and a bit $C \in \set{0,1}$ from Bob, $\F_\text{12ROT}$ sends $S_C$ to Bob.
\vspace{-1ex}
\end{framed}
\vspace{-2ex}
\small
\caption{The ideal Randomized 1-2 OT functionality.}\label{fig:OT-func}
\vspace{-1ex}
\end{figure}

We recall the security definition of randomized 1-2 OT
from~\cite{DFRSS07}.  The definition is in the framework described in
Section~\ref{sec:SecDef} and considers a single execution of the
protocol with an arbitrary distribution for honest Bob's input bit and
no input for the dishonest players. For consistency with common
notation, we denote Bob's input $V$ by $C$ (whereas Alice input is
empty), and Alice's outputs by $X = (S_0,S_1)$.

\begin{definition}[\RandlStringOT] \label{def:Rl12OT}
A randomized 1-2 OT protocol is {\em $\eps$-secure} (against $\mathfrak{A}$ and $\mathfrak{B}$) if the following properties hold. 
\begin{description}
\item[{\em Correctness:}] If Alice and Bob are honest, then for any distribution of Bob's input $C$, $S_0$ and $S_1$ are $\eps$-close to random and independent of $C$, and $Y = S_C$ except with probability~$\eps$.
\item[{\em Security for Alice:}] 
For any dishonest $\dB \in \mathfrak{B}$, the resulting common output state $\rho_{S_0 S_1 Y'}$ allows a classical binary $C$ such
  that $\rho_{S_{1-C} S_{C} C Y'} \epsclose \frac{1}{2^\ell}\I \otimes \rho_{S_{C} C Y'}$.
\item[{\em Security for Bob:}]  
For any dishonest $\dA \in \mathfrak{A}$, and for any distribution of $C$, the resulting common output state $\rho_{X' C Y}$ (augmented with $C$) allows classical $S_0,S_1$ such that \smash{$\Pr\bigl[Y=S_C\bigr] \geq 1-\varepsilon$} and $\rho_{S_0 S_1 X' C}  \epsclose \rho_{S_0 S_1 X'} \otimes \rho_C$.
\end{description}
\end{definition}
Note that the correctness condition in Definition~\ref{def:Rl12OT} is
somewhat stronger than the correctness condition in the
definition proposed in~\cite{DFRSS07}, which merely requires that $Y =
S_C$ except with probability~$\eps$. We point out that this difference
is not crucial for Proposition~\ref{prop:OTEx} below to hold. Indeed, if $Y =
S_C$ is guaranteed with high probability, then correctness as
in~Definition~\ref{def:Rl12OT} can be bootstrapped from the security
properties for {\em dishonest} players, albeit with some loss in the error
probability: security for Bob guarantees that the distribution of
$(S_0,S_1)$ is close to independent of $C$, and security for
Alice guarantees that the distribution of $S_0$, which is close to the
distribution of $S_0$ conditioned on $C=1$, is random and independent
of $S_1$ (conditioned on $C=1$ or not), and similar for $S_1$. Working
out the details is tedious\footnote{What makes it particularly tedious
  is that e.g.\ the random variable $C$ that is guaranteed to exist by
  the security for Alice may a-priori differ from honest Bob's $C$,
  and one has to explicitly argue that they have to be close. } and
does not give any new insight. In most circumstances, such an argument
is not even needed. For any given protocol, the correctness condition of
Definition~\ref{def:security} can typically be trivially verified by
inspection.

\begin{proposition}\label{prop:OTEx}
  A quantum protocol satisfying Definition~\ref{def:Rl12OT}
  $4\eps$-securely implements $\F_\text{12ROT}$ according to
  Definition~\ref{def:security}. 
\end{proposition}

\begin{proof}
Correctness follows immediately.

{\em Security for Alice:} 
Consider $C$ which is guaranteed to exist by Definition~\ref{def:Rl12OT}. Let us define $V = (C,S_C)$ and $Y = \emp$. As Alice's input $U$ is empty, $V$ is trivially independent of $U$. 
Note that, $\rho_{S_{1-C} S_{C} C Y'} \epsclose 2^{-\ell}\I \otimes
\rho_{S_{C} C Y'}$ in particular implies that $P_{S_{1-C} S_{C} C}
\epsclose 2^{-\ell} P_{S_{C} C}$. Therefore it follows that
$$
\big(\emp,(C,S_C),(S_0,S_1), \emp\big) \epsclose \big(\emp,(C,S_C),\F_\text{12ROT}(\emp,(C,S_C))\big)
$$
by the definition of $\F_\text{12ROT}$. Finally, by the third claim of
Lemma~\ref{lemma:right->mid}, $\rho_{S_{1-C} S_{C} C Y'} \epsclose 2^{-\ell}\I \otimes \rho_{S_{C} C Y'}$ implies that
$$
\rho_{S_{1-C} S_C C Y'} \epsclose[4\eps] \rho_{\MC{S_{1-C}}{S_C C}{Y'}} 
$$ 
from which it follows, by Lemma~\ref{lemma:extend}, that
$$
\rho_{\emp(S_0 S_1)(C S_C)\emp Y'}  \epsclose[4\eps] \rho_{\MC{\emp(S_0 S_1)}{(C S_C)\emp}{Y'}} \, .
$$

{\em Security for Bob:} 
Consider $S_0,S_1$ which is guaranteed to exist by Definition~\ref{def:Rl12OT}. Let us define $U = (S_0,S_1)$ and $X = \emp$.  $\rho_{S_0 S_1 X' C}  \epsclose \rho_{S_0 S_1 X'} \otimes \rho_C$ in particular implies that $P_{S_0 S_1 C} \epsclose P_{S_0 S_1} P_C$. 
Furthermore,  it is easy to see that $P[Y\!\neq\! S_C] \leq \eps$ implies 
$$
\big((S_0,S_1),C,\emp,Y\big) \epsclose \big((S_0,S_1),C,\emp,S_C\big) = \big((S_0,S_1),C,\F_\text{12ROT}((S_0,S_1),C)\big) \, .
$$
Finally, by Lemma~\ref{lemma:right->mid}, $\rho_{S_0 S_1 X' C}  \epsclose \rho_{S_0 S_1 X'} \otimes \rho_C$ in particular implies $\rho_{C S_0 S_1 X'} \epsclose[2\eps] \rho_{\MC{C}{S_0 S_1}{X'}}$, from which follows by Lemma~\ref{lemma:extend} that $\rho_{C S_C S_0 S_1 X'} \epsclose[2\eps] \rho_{\MC{C S_C}{S_0 S_1}{X'}}$. Using \mbox{$P[Y\!\neq\! S_C] \leq \eps$}, this implies 
$$
\rho_{C Y (S_0 S_1)\emp X'} \epsclose[3\eps] \rho_{\MC{C Y}{(S_0 S_1) \emp}{X'}} \, .
$$
The last claim follows from the following observation. 
\begin{align*}
\delta\big(\rho_{C Y S_0 S_1 X'}&, \rho_{\MC{C Y}{S_0 S_1}{X'}} \big) = \sum_{c y s_0 s_1} P_{C Y S_0 S_1}(c,y,s_0,s_1) \, \delta(\rho_{X'}^{c y s_0 s_1},\rho_{X'}^{s_0 s_1}) \\
&=  P[Y\!=\!S_C] \cdot \!\sum_{c y s_0 s_1} P_{C Y S_0 S_1|Y=S_C}(c,y,s_0,s_1) \, \delta(\rho_{X'}^{c y s_0 s_1},\rho_{X'}^{s_0 s_1}) + P[Y\!\neq\!S_C] \cdot \textit{rest}\\
&=  P[Y\!=\!S_C] \cdot \!\sum_{c s_0 s_1} P_{C S_0 S_1|Y=S_C}(c,s_0,s_1) \, \delta(\rho_{X'|Y=S_C}^{c s_0 s_1},\rho_{X'}^{s_0 s_1}) + P[Y\!\neq\!S_C] \cdot \textit{rest}
\end{align*}
where $0 \leq \textit{rest} \leq 1$, and similarly for $\delta\big(\rho_{C S_C
  S_0 S_1 X'}, \rho_{\MC{C S_C}{S_0 S_1}{X'}} \big)$. Subtracting the
two terms results in a value that is upper bounded by $P[Y\!\neq\!S_C] \leq \eps$ in absolute value. 
\qed
\end{proof}

\section{Conclusion}

We proposed a general security definition for quantum protocols in
terms of simple quantum-information-theoretic conditions and showed
that quantum protocols fulfilling the definition do their job as
expected when used as subroutines in a larger classical protocol.
The restriction to classical ``outer'' protocols fits our currently
limited ability for executing quantum protocols, but can also be
appreciated in that our security conditions pose the {\em minimal}
requirements for a quantum protocol to be useful beyond running it in
isolation.

\section*{Acknowledgements}
We would like to thank J\"urg Wullschleger for sharing a draft
of~\cite{CW08} and pointing out how to avoid the dependency from
the input distribution of the dishonest player in the ideal model.


\bibliographystyle{alpha} 
\bibliography{qip,crypto,procs}

\newcommand{\etalchar}[1]{$^{#1}$}
\begin{thebibliography}{KRBM07}

\bibitem[BCG{\etalchar{+}}05]{BCGHS05}
Michael {Ben-Or}, Claude Cr{\'e}peau, Daniel Gottesman, Avinatan Hassidim, and
  Adam Smith.
\newblock Secure multiparty quantum computation with (only) a strict honest
  majority.
\newblock In {\em 46th Annual IEEE Symposium on Foundations of Computer Science
  (FOCS)}, pages 249--260, 2005.

\bibitem[BHL{\etalchar{+}}05]{BHLMO05}
Michael {Ben-Or}, Michal Horodecki, Debbie~W. Leung, Dominic Mayers, and
  Jonathan Oppenheim.
\newblock The universal composable security of quantum key distribution.
\newblock In {\em Theory of Cryptography Conference (TCC)}, volume 3378 of {\em
  Lecture Notes in Computer Science}, pages 386--406. Springer, 2005.

\bibitem[BM04]{BM04}
Michael {Ben-Or} and Dominic Mayers.
\newblock General security definition and composability for quantum and
  classical protocols, September 2004.
\newblock {\tt http://arxive.org/abs/quant-ph/0409062}.

\bibitem[Cr{\'e}87]{Crepeau87}
Claude Cr{\'e}peau.
\newblock Equivalence between two flavours of oblivious transfers.
\newblock In {\em Advances in Cryptology---CRYPTO~'87}, volume 293 of {\em
  Lecture Notes in Computer Science}. Springer, 1987.

\bibitem[CSSW06]{CSSW06}
Claude Cr{\'e}peau, George Savvides, Christian Schaffner, and J{\"u}rg
  Wullschleger.
\newblock Information-theoretic conditions for two-party secure function
  evaluation.
\newblock In {\em Advances in Cryptology---EUROCRYPT~'06}, volume 4004 of {\em
  Lecture Notes in Computer Science}, pages 538--554. Springer, 2006.

\bibitem[CW08]{CW08}
Claude Cr{\'e}peau and J{\"u}rg Wullschleger.
\newblock Statistical security conditions for two-party secure function
  evaluation.
\newblock In {\em Third International Conference on Information Theoretic
  Security (ICITS)}, pages 86--99, 2008.

\bibitem[DFR{\etalchar{+}}07]{DFRSS07}
Ivan~B. Damg{\aa}rd, Serge Fehr, Renato Renner, Louis Salvail, and Christian
  Schaffner.
\newblock A tight high-order entropic quantum uncertainty relation with
  applications.
\newblock In {\em Advances in Cryptology---CRYPTO~'07}, volume 4622 of {\em
  Lecture Notes in Computer Science}, pages 360--378. Springer, 2007.

\bibitem[DFSS05]{DFSS05}
Ivan~B. Damg{\aa}rd, Serge Fehr, Louis Salvail, and Christian Schaffner.
\newblock Cryptography in the bounded quantum-storage model.
\newblock In {\em 46th Annual IEEE Symposium on Foundations of Computer Science
  (FOCS)}, pages 449--458, 2005.
\newblock Full version available at: {\tt
  http://arxiv.org/abs/quant-ph/0508222v2}.

\bibitem[DFSS07]{DFSS07}
Ivan~B. Damg{\aa}rd, Serge Fehr, Louis Salvail, and Christian Schaffner.
\newblock Secure identification and {QKD} in the bounded-quantum-storage model.
\newblock In {\em Advances in Cryptology---CRYPTO~'07}, volume 4622 of {\em
  Lecture Notes in Computer Science}, pages 342--359. Springer, 2007.

\bibitem[DFSS08]{DFSS08}
Ivan~B. Damg{\aa}rd, Serge Fehr, Louis Salvail, and Christian Schaffner.
\newblock Cryptography in the bounded-quantum-storage model.
\newblock {\em SIAM Journal on Computing}, 37(6):1865--1890, 2008.

\bibitem[GL03]{GotLo03}
Daniel Gottesman and Hoi-Kwong Lo.
\newblock Proof of security of quantum key distribution with two-way classical
  communications.
\newblock {\em IEEE Transactions on Information Theory}, 49(2):457--475, 2003.
\newblock {\tt http://arxiv.org/abs/quant-ph/0105121}.

\bibitem[Gol04]{Goldreich04}
Oded Goldreich.
\newblock {\em Foundations of Cryptography}, volume II: Basic Applications.
\newblock Cambridge University Press, 2004.

\bibitem[Kil88]{Kilian88}
Joe Kilian.
\newblock Founding cryptography on oblivious transfer.
\newblock In {\em 20th Annual ACM Symposium on Theory of Computing (STOC)},
  pages 20--31, 1988.

\bibitem[Kit03]{Kitaev03}
A.~Kitaev.
\newblock Quantum coin-flipping.
\newblock presented at QIP'03. A review of this technique can be found in
  {\tt\verb+http://lightlike.com/~carlosm/publ+}, 2003.

\bibitem[KRBM07]{KRBM07}
Robert Koenig, Renato Renner, Andor Bariska, and Ueli Maurer.
\newblock Small accessible quantum information does not imply security.
\newblock {\em Physical Review Letters}, 98(140502), April 2007.

\bibitem[LC97]{LC97}
Hong-Kwong Lo and H.~F. Chau.
\newblock Is quantum bit commitment really possible?
\newblock {\em Physical Review Letters}, 78(17):3410--3413, April 1997.

\bibitem[Lo97]{Lo97}
Hong-Kwong Lo.
\newblock Insecurity of quantum secure computations.
\newblock {\em Physical Review A}, 56(2):1154--1162, 1997.

\bibitem[May97]{Mayers97}
Dominic Mayers.
\newblock Unconditionally secure quantum bit commitment is impossible.
\newblock {\em Physical Review Letters}, 78(17):3414--3417, April 1997.

\bibitem[NC00]{NC00}
Michael~A. Nielsen and Isaac~L. Chuang.
\newblock {\em Quantum Computation and Quantum Information}.
\newblock Cambridge university press, 2000.

\bibitem[Ren05]{Renner05}
Renato Renner.
\newblock {\em Security of Quantum Key Distribution}.
\newblock PhD thesis, ETH Z\"urich (Switzerland), September 2005.
\newblock {\tt http://arxiv.org/abs/quant-ph/0512258}.

\bibitem[RK05]{RK05}
Renato Renner and Robert K\"onig.
\newblock Universally composable privacy amplification against quantum
  adversaries.
\newblock In {\em Theory of Cryptography Conference (TCC)}, volume 3378 of {\em
  Lecture Notes in Computer Science}, pages 407--425. Springer, 2005.

\bibitem[Sal98]{Salvail98}
Louis Salvail.
\newblock Quantum bit commitment from a physical assumption.
\newblock In {\em Advances in Cryptology---CRYPTO~'98}, volume 1462 of {\em
  Lecture Notes in Computer Science}, pages 338--353. Springer, 1998.

\bibitem[Unr04]{Unruh04}
Dominique Unruh.
\newblock Simulatable security for quantum protocols.
\newblock {\tt http://arxiv.org/abs/quant-ph/0409125}, 2004.

\bibitem[WST07]{WST07arxiv}
Stephanie Wehner, Christian Schaffner, and Barbara Terhal.
\newblock Practical cryptography from noisy photonic storage.
\newblock {\tt http://arxiv.org/abs/0711.2895}, 2007.

\bibitem[WW08]{WW08}
Stephanie Wehner and J{\"u}rg Wullschleger.
\newblock Composable security in the bounded-quantum-storage model.
\newblock In {\em Automata, Languages and Programming, 35th International
  Colloquium, ICALP 2008}, volume 5126 of {\em Lecture Notes in Computer
  Science}, pages 604--615. Springer, 2008.

\end{thebibliography}

\appendix

\section{Proofs}\label{app:proofs}

\subsection{Proof of Lemma~\ref{lemma:right->mid}}

We show that $\rho_{\MC{X}{Y}{ZE}} \epsclose \rho_{\MC{X}{YZ}{E}}$, the first claim then follows by triangle inequality. 
Since quantum operations do not increase the trace distance, tracing out register $E$ in $\rho_{XYZE}$ and $\rho'_{XYZE} \assign \rho_{\MC{X}{Y}{ZE}}$ implies that 
$$
P_{XYZ} \approx_\eps P_{\MC{X}{Y}{Z}} = P_{XY} \cdot P_{Z|Y} \, .
$$
By elementary properties of the trace distance, it follows that
\begin{align*}
\delta\bigl( \rho&_{\MC{X}{Y}{ZE}}, \rho_{\MC{X}{YZ}{E}} \bigr) 
= \sum_{x,y,z} P_{XY}(x,y) \delta\bigl(P_{Z|Y}(z|y)\rho_E^{y,z}, P_{Z|XY}(z|x,y) \rho_E^{y,z} \bigr) \\
&= \frac12 \sum_{x,y,z} P_{XY}(x,y) \big| P_{Z|Y}(z|y) - P_{Z|XY}(z|x,y) \big| \tr\bigl| \rho_E^{y,z} \bigr| 
= \delta\bigl( P_{\MC{X}{Y}{Z}}, P_{XYZ} \bigr) \leq \eps \, .
\end{align*}

The second claim follows by letting $Y$ be ``empty".  The third claim
holds because by the triangle inequality, we have $$\delta(\rho_{XZE},
\rho_X \otimes \rho_{ZE}) \leq \delta(\rho_{XZE}, \I/|{\cal X}|
\otimes \rho_{ZE}) + \delta(\I/|{\cal X}| \otimes \rho_{ZE}, \rho_X
\otimes \rho_{ZE}) \leq 2 \eps$$ and we can then use the second claim.
\qed

\subsection{Proof of Lemma~\ref{lemma:extend}}

By elementary properties of the trace distance, 
\begin{align*}
\delta\bigl(\rho_{Xf(X,Y)YE}, & \,\rho_{\MC{Xf(X,Y)}{Y}{E}} \bigr) 
=  \sum_{x,z,y} P_{X f(X,Y) Y}(x,z,y) \delta\bigl( \rho_E^{x,y}, \rho_E^{y} \bigr) \\
 &=  \sum_{x,y} P_{X Y}(x,y) \delta\bigl( \rho_E^{x,y}, \rho_E^{y} \bigr) 
= \delta(\rho_{XYE},\rho_{\MC{X}{Y}{E}} ) \leq \eps\, .
\end{align*}
\qed

\subsection{Proof of Lemma~\ref{lemma:event}}
Let $p=\Pr[\ev]$ and $\ol{p}=\Pr[\ol{\ev}]$ and define the two sets
$\cY_\ev = \Set{y}{\Pr[\ev|Y=y]=1}$ and $\cY_{\ol{\ev}} =
\Set{y}{\Pr[\ol{\ev}|Y=y]=1}$. Then,
\begin{align*}
\rho_{\MC{X}{Y}{Z}} &= \sum_{x,y} P_{XY}(x,y) \proj{x} \otimes
\proj{y} \otimes \rho_E^{y} \\
&= \sum_{x,y\in \cY_\ev} P_{XY}(x,y) \proj{x} \otimes
\proj{y} \otimes \rho_{E}^{y} + \sum_{x,y\in \cY_{\ol{\ev}}} P_{XY}(x,y) \proj{x} \otimes
\proj{y} \otimes \rho_{E}^{y}\\
&= \sum_{x,y\in \cY_\ev} p \cdot P_{XY|\ev}(x,y) \proj{x} \otimes
\proj{y} \otimes \rho_{E|\ev}^{y} + \sum_{x,y\in \cY_{\ol{\ev}}} \ol{p}
\cdot P_{XY|\ol{\ev}}(x,y) \proj{x} \otimes
\proj{y} \otimes \rho_{E|\ol{\ev}}^{y}\\
&= p \cdot \rho_{\MC{X}{Y}{E}|\ev} +
\ol{p} \cdot \rho_{\MC{X}{Y}{E}|\overline{\ev}} \, ,
\end{align*}
where we used in the third equality that for $y \in \cY_\ev$, it
follows from the assumption over the event that $\rho_E^y =
\rho_{E|\ev}^y$ and similarly for $y \in \cY_{\ol{\ev}}$.
\qed

\section{Other Variants of Oblivious Transfer}\label{app:otherOT}
In this section, we give analogous ``minimal'' requirements for
composability of other variants of oblivious transfer. It has been
shown~\cite{Crepeau87} that all these variants of oblivious transfer
are equivalent and universal for secure two-party function
evaluation~\cite{Kilian88}. In fact, the results of this paper show
that if the variants are implemented by a quantum protocol according
to our security definitions, these classical results still hold.

\subsection{Regular (Non-Randomized) 1-2 OT}
Figure~\ref{fig:OT-funcStandard} shows the ideal functionality for the
standard (non-randomized) 1-out-of-2 String-OT. It takes two input strings
$S_0$ and $S_1$ of $\ell$ bits each from Alice and an input bit $C$
from Bob, and it outputs an $\ell$-bit string $Y$ which stands for the
string of his choice $S_C$ to Bob.
\begin{figure}
\normalsize
\begin{framed}
{\bf Functionality} $\F_\text{12OT}$:  
Upon receiving $S_0, S_1 \in \set{0,1}^\ell$ from Alice and a choice bit $C \in
\set{0,1}$ from Bob, $\F_\text{12OT}$ sends $S_C$ to Bob.
\end{framed}
\vspace{-2ex}
\small
\caption{The ideal 1-2 OT functionality.}\label{fig:OT-funcStandard}
\end{figure}

The definition is in the framework described in
Section~\ref{sec:SecDef} and considers a single execution of the
protocol with an arbitrary distribution for honest Bob's input bit and
no input for the dishonest players. For consistency with common
notation, we denote Alice's input $U$ by $(S_0,S_1)$ and Bob's input
$V$ by $C$.

\begin{definition}[\lStringOT] \label{def:l12OT}
A 1-2 OT protocol is {\em $\eps$-secure} if the following properties hold. 
\begin{description}
\item[{\em Correctness:}] If Alice and Bob are honest, then for any
  joint distribution of Alice's inputs $S_0,S_1$ and Bob's input $C$, it
  holds that Bob's output $Y = S_C$ except with probability~$\eps$.
\item[{\em Security for Alice:}] If Alice is honest, then for any
  dishonest Bob and any distribution of Alice's inputs $S_0,S_1$,
  Alice does not get any output and the
  common output state $\rho_{S_0 S_1 Y'}$ allows a classical binary
  $C$ such that $\rho_{S_0 S_1 C} \epsclose \rho_{S_0 S_1} \otimes
  \rho_C$ and $\rho_{S_{1-C} S_{C} C Y'} \epsclose
  \rho_{\MC{S_{1-C}}{S_C C}{Y'}}$.
\item[{\em Security for Bob:}] If Bob is honest, then for any
  dishonest Alice and any distribution of Bob's input $C$, the common
  output state $\rho_{X' C Y}$ allows classical $S_0,S_1$ such that
  \smash{$\Pr\bigl[Y=S_C\bigr] \geq 1-\varepsilon$} and $\rho_{S_0 S_1
    X' C} \epsclose \rho_{S_0 S_1 X'} \otimes \rho_C$.
\end{description}
\end{definition}

\begin{proposition}\label{prop:RegularOTEx}
  A quantum protocol satisfying Definition~\ref{def:l12OT}
  $3\eps$-securely implements $\F_\text{12OT}$ according to
  Definition~\ref{def:security}. 
\end{proposition}

\begin{proof} Correctness follows immediately.

{\em Security for Alice:} 
Consider $C$ which is guaranteed to exist by
Definition~\ref{def:l12OT}. 
Let us define $V = C$ and $Y = S_C$. By the first requirement in the
definition, we have
that Alice's input $S_0,S_1$ is $\eps$-close to independent of $C$.

Furthermore, it holds by definition that
\[((S_0,S_1),C,\emp,S_C) =
((S_0,S_1),C,\F_{\text{12OT}}((S_0,S_1),C)) \, .\]

Finally, by the second requirement in the definition and 
Lemma~\ref{lemma:extend}, we have that
\[ \rho_{(S_0 S_1) C S_C Y'} \epsclose \rho_{\MC{S_0 S_1}{C S_C}{Y'}}
\, .
\]

{\em Security for Bob:} as in the proof of Proposition~\ref{prop:OTEx}.
\qed
\end{proof}

\subsection{Fully Randomized 1-2 OT}
Figure~\ref{fig:OK-func} below shows the ideal functionality for
fully randomized 1-2 String-OT (sometimes also called Oblivious Key OK). It
takes no input from the players and outputs two random $\ell$-bit
strings $S_0$ and $S_1$ to Alice, a random choice bit $C$ and $S_C$ to
Bob. Note that it allows a dishonest Alice to influence the
distribution of $S_0$ and $S_1$, and a dishonest Bob to influence the
distribution of $S_C$; but this is good enough for many applications,
in particular to build a regular (non-randomized) 1-2 OT in the
standard manner.

\begin{figure}
\normalsize
\begin{framed}
{\bf Functionality} $\F_\text{12OK}$: 

\medskip{\em Honestly behaving Alice and Bob}: Upon receiving no input
from Alice and Bob, $\F_\text{12OK}$ samples two random and
independent strings $S_0,S_1 \in_R \set{0,1}^\ell$ and a choice bit $C
\in_R \set{0,1}$, and sends $S_0,S_1$ to Alice and $C,S_C$ to Bob.

\medskip{\em Honest Alice and dishonest Bob}: Upon receiving no input
from Alice and a bit $C \in \set{0,1}$ and a string $S_C \in
\set{0,1}^\ell$ from Bob, $\F_\text{12OK}$ samples a random
independent string $S_{1-C} \in_R \set{0,1}^\ell$, and sends $S_0$ and
$S_1$ to Alice.

\medskip{\em Dishonest Alice and honest Bob}: Upon receiving two
strings $S_0,S_1 \in \set{0,1}^\ell$ from Alice and no input from Bob, $\F_\text{12OK}$
samples a random bit $C \in_R \set{0,1}$ and sends $C, S_C$ to Bob.
\vspace{-1ex}
\end{framed}
\vspace{-2ex}
\small
\caption{The ideal Randomized 1-2 OT functionality.}\label{fig:OK-func}
\end{figure}

The following definition is in the framework described in
Section~\ref{sec:SecDef} and considers a single execution of the
protocol with no inputs for honest or dishonest players. For
consistency with common notation, we denote Alice's output $X$ by
$(S_0,S_1)$ and Bob's output by $(C, Y)$.

\begin{definition}[Fully Randomized \lStringOT] \label{def:12OK}
A randomized 1-2 OT protocol is {\em $\eps$-secure} if the following properties hold. 
\begin{description}
\item[{\em Correctness:}] If Alice and Bob are honest, then $S_0, S_1$
  and $C$ are $\eps$-close to random and independent, and $Y = S_C$
  except with probability~$\eps$.
\item[{\em Security for Alice:}] If Alice is honest, then for any
  dishonest Bob, the common output state $\rho_{S_0 S_1 Y'}$ allows a
  classical binary $C$ such that $\rho_{S_{1-C} S_{C} C Y'} \epsclose
  \frac{1}{2^\ell}\I \otimes \rho_{S_{C} C Y'}$.
\item[{\em Security for Bob:}] If Bob is honest, then for any
  dishonest Alice, the common output state $\rho_{X' C Y}$ allows
  classical $S_0,S_1$ such that \smash{$\Pr\bigl[Y=S_C\bigr] \geq
    1-\varepsilon$} and $\rho_{S_0 S_1 X' C} \epsclose \rho_{S_0 S_1
    X'} \otimes \I/2$.
\end{description}
\end{definition}

\begin{proposition}\label{prop:OKEx}
  A quantum protocol satisfying Definition~\ref{def:12OK}
  $4\eps$-securely implements $\F_\text{12OK}$ according to
  Definition~\ref{def:security}. 
\end{proposition}

\begin{proof}
Correctness follows immediately.

{\em Security for Alice:} as in Proposition~\ref{prop:OTEx}.

{\em Security for Bob:} Consider $S_0,S_1$ which is guaranteed to
exist by Definition~\ref{def:Rl12OT}. Let us define Alice's input $U =
(S_0,S_1)$ and Alice's output $X = \emp$. The requirement $\rho_{S_0 S_1 X' C}
\epsclose \rho_{S_0 S_1 X'} \otimes \I/2$ in particular implies that
$P_{S_0 S_1 C} \epsclose P_{S_0 S_1} P_U$.  Furthermore, it is easy to
see that $P[Y\!\neq\! S_C] \leq \eps$ implies
$$
\big((S_0,S_1),\emp,\emp,(C,Y)\big) \epsclose \big((S_0,S_1),\emp,\emp,(C,S_C)\big) \epsclose \big((S_0,S_1),\emp,\F_\text{12ROT}((S_0,S_1),\emp)\big) \, .
$$
Finally, by Lemma~\ref{lemma:right->mid}, $\rho_{S_0 S_1 X' C}
\epsclose \rho_{S_0 S_1 X'} \otimes \I/2$ implies
$\rho_{C S_0 S_1 X'} \epsclose[4\eps] \rho_{\MC{C}{S_0 S_1}{X'}}$,
from which follows by Lemma~\ref{lemma:extend} that $\rho_{C S_C S_0
  S_1 X'} \epsclose[8\eps] \rho_{\MC{C S_C}{S_0 S_1}{X'}}$. Using
\mbox{$P[Y\!\neq\! S_C] \leq \eps$}, this implies
$$
\rho_{\emp (C Y) (S_0 S_1)\emp X'} \epsclose[9\eps] \rho_{\MC{\emp (C Y)}{(S_0 S_1) \emp}{X'}} \, .
$$
The last claim follows from the following observation. 
\begin{align*}
\delta\big(&\rho_{C Y S_0 S_1 X'}, \rho_{\MC{C Y}{S_0 S_1}{X'}} \big) = \sum_{c y s_0 s_1} P_{C Y S_0 S_1}(c,y,s_0,s_1) \, \delta(\rho_{X'}^{c y s_0 s_1},\rho_{X'}^{s_0 s_1}) \\
&=  P[Y\!=\!S_C] \cdot \!\sum_{c y s_0 s_1} P_{C Y S_0 S_1|Y=S_C}(c,y,s_0,s_1) \, \delta(\rho_{X'}^{c y s_0 s_1},\rho_{X'}^{s_0 s_1}) + P[Y\!\neq\!S_C] \cdot \textit{rest}\\
&=  P[Y\!=\!S_C] \cdot \!\sum_{c s_0 s_1} P_{C S_0 S_1|Y=S_C}(c,s_0,s_1) \, \delta(\rho_{X'|Y=S_C}^{c s_0 s_1},\rho_{X'}^{s_0 s_1}) + P[Y\!\neq\!S_C] \cdot \textit{rest}
\end{align*}
where $0 \leq \textit{rest} \leq 1$, and similarly for $\delta\big(\rho_{C S_C
  S_0 S_1 X'}, \rho_{\MC{C S_C}{S_0 S_1}{X'}} \big)$. Subtracting the
two terms results in a value that is upper bounded by $P[Y\!\neq\!S_C] \leq \eps$ in absolute value. 
\qed
\end{proof}

\subsection{Randomized Rabin OT}
Figure~\ref{fig:RabinOT-func} shows the ideal functionality for
(randomized) Rabin Oblivious Transfer. It samples a uniform random bit
$C \in_R \set{0,1}$ and a string $S \in_R \set{0,1}^\ell$. It outputs
$S$ to Alice, $C$ to Bob and in case $C=1$, also $S$ is output to
Bob. If $C=0$, Bob receives the all-0 string.

\medskip
\begin{figure}
\normalsize
\begin{framed}
{\bf Functionality} $\F_\text{RabinOT}$:  

\medskip{\em Honestly behaving Alice and Bob}: 
Upon receiving no input from the players, $\F_\text{RabinOT}$
samples $S \in \set{0,1}^\ell$ and $C \in_R \set{0,1}$ and sends $X
\assign S$ to Alice and $C, Y \assign C \cdot S$ to Bob.

\medskip{\em Honest Alice and dishonest Bob}: 
Upon receiving no input from Alice and a string $S \in \set{0,1}^\ell$
from Bob, $\F_\text{RabinOT}$ samples a random independent bit $C$ and
outputs it to Bob. If $C=1$, $\F_\text{RabinOT}$ sends $X=S$ to Alice. If
$C=0$, $\F_\text{RabinOT}$ samples a new string $S' \in_R
\set{0,1}^\ell$ and sends $X=S'$ to Alice.

\medskip{\em Dishonest Alice and honest Bob}: 
Upon receiving a string $S \in \set{0,1}^\ell$ from Alice and no input
from Bob, $\F_\text{RabinOT}$ samples a bit $C \in_R \set{0,1}$ and
sends $C, C \cdot S$ to Bob and no output to Alice.

\end{framed}
\vspace{-2ex}
\small
\caption{The ideal Rabin OT functionality.}\label{fig:RabinOT-func}
\end{figure}
\medskip

The following definition is in the framework described in
Section~\ref{sec:SecDef} and considers a single execution of the
protocol with no inputs for honest or dishonest players. For
consistency with common notation, we denote Bob's output by $(C, Y)$.

\begin{definition}[Rabin OT] \label{def:RabinOT} A randomized Rabin-OT
  protocol is {\em $\eps$-secure} if the following properties hold.
\begin{description}
\item[{\em Correctness:}] If Alice and Bob are honest, then $X$ and
  $C$ are $\eps$-close to random and independent and $Y = C \cdot X$
  except with probability~$\eps$.
\item[{\em Security for Alice:}] If Alice is honest, then for any
  dishonest Bob, the common output state $\rho_{X Y'}$ allows a classical binary
  $C$ such that $\rho_{X C} \epsclose \rho_{X} \otimes
  \I/2$ and $\rho_{X C Y'|C=0} \epsclose \I/2^\ell \otimes \rho_{C Y'|C=0}$.
\item[{\em Security for Bob:}] If Bob is honest, then for any
  dishonest Alice, the common output state $\rho_{X' C Y}$ allows a
  classical $S$ such that $Y=C \cdot S$ except with probability $\eps$
  and $\rho_{C S X'} \epsclose \I/2 \otimes \rho_{S X'}$.
\end{description}
\end{definition}

\begin{proposition}\label{prop:RabinOTEx}
  A quantum protocol satisfying Definition~\ref{def:RabinOT}
  $5\eps$-securely implements $\F_\mathrm{RabinOT}$ according to
  Definition~\ref{def:security}. 
\end{proposition}

\begin{proof} Correctness follows immediately.

{\em Security for Alice:} 
Consider $C$ which is guaranteed to exist by
Definition~\ref{def:RabinOT}.  
Let us define Bob's input $V \assign X$ if $C=1$. In case $C=0$, sample $V
\in_R \set{0,1}^\ell$. Let Bob's output be
$Y = (C,C \cdot V)$. As Alice has no input $U=\emp$, Bob's input $V$ is
trivially independent of $U$.  Furthermore, the definition requires
$C$ to be $\eps$-close to independent from $X$ and to completely
random. In case $C=1$, $\F_{\text{RabinOT}}$ outputs $(1,Y)=(1,1 \cdot
V)=(1,1 \cdot X)$ to
Bob and $X$ to Alice. In case $C=0$, Bob receives $(0,0)$ and Alice's
output $X$ is independent of Bob's input $V$. Hence,
\[(\emp, V, X,(C, C \cdot V)) \epsclose
(\emp,V,\F_{\text{RabinOT}}(\emp,V)) \, .\]

From $\rho_{X C Y'|C=0} \epsclose \I/2^\ell \otimes \rho_{C Y'|C=0}$
follows by third claim of Lemma~\ref{lemma:right->mid} that 
$\rho_{X C Y'|C=0} \epsclose[4\eps] \rho_{\MC{X}{C}{Y'}|C=0}$, and as
$V$ is sampled at random, also $\rho_{X V C Y'|C=0} \epsclose[4\eps]
\rho_{\MC{X}{V C}{Y'}|C=0}$ holds. It follows that
\begin{align*}
  \rho_{X V (C, C \cdot V) Y'} &= \Pr[C=0] \cdot \rho_{X V (C, C \cdot V)
    Y'|C=0} + \Pr[C=1] \cdot \rho_{X V (C, C \cdot V)
    Y'|C=1} \\
  &= \Pr[C=0] \cdot \rho_{X V (C, C \cdot V) Y'|C=0} + \Pr[C=1] \cdot
  \rho_{\MC{X}{V (C, C \cdot V)}{Y'}|C=1}\\
  &\epsclose[4\eps] \Pr[C=0] \cdot \rho_{\MC{X}{V (C, C \cdot V)}{Y'}|C=0} + \Pr[C=1] \cdot
  \rho_{\MC{X}{V (C, C \cdot V)}{Y'}|C=1}\\
  &\epsclose \rho_{\MC{X}{V (C, C \cdot V)}{Y'}} \, ,
\end{align*}
where we used Lemma~\ref{lemma:event} for the last approximation.

{\em Security for Bob:} Consider $S$ which is guaranteed to exist by
Definition~\ref{def:RabinOT}. Let us define Alice's input to be $U
\assign S$ and let Alice's output $X$ be empty. As Bob does not have
input $V=\emp$, Alice's $U$ is trivially independent of
$V$. Furthermore, since $C$ is $\eps$-close to uniformly random and
independent of $S$ and $Y=C \cdot S$ except with probability $\eps$,
we have
\[ (S,\emp,\emp,(C,Y)) \epsclose[2\eps]
(S,\emp,\F_{\text{RabinOT}}(S,\emp)) \, .
\]

From $\rho_{C S X'} \epsclose \I/2 \otimes \rho_{S X'}$ follows that
$\rho_{C S X'} \epsclose[2\eps] \rho_C \otimes \rho_{S X'}$ and
therefore by Lemma~\ref{lemma:right->mid}, $\rho_{C S X'}
\epsclose[4\eps] \rho_{\MC{C}{S}{X'}}$. As $Y=C \cdot S$ except with
probability $\eps$, we have by Lemma~\ref{lemma:extend} that
\[ \rho_{\emp C Y S \emp X'} \epsclose[5\eps] \rho_{\MC{C Y}{S}{X'}}
\, .
\]
\qed
\end{proof}

\section{Secure Identification without Unfairness}\label{sec:identsimpler}

The goal of this section is to provide a slightly stronger
functionality for secure identification than the one presented in
Section~\ref{sec:ident}. It is stronger in that we do not allow
dishonest Alice to make Bob reject {\em while she learns whether $W_A
  = W_B$ or not}, but we still allow Alice to make Bob reject all the
time by inputting a symbol $\perp$ that never agrees with Bob's input, see
Figure~\ref{fig:identfuncsimpler}.  In order to achieve this
functionality, we have to impose a slightly stricter security
definition than Definition~\ref{def:ident}.

\begin{figure}
\normalsize
\begin{framed} {\bf Functionality} $\F_{\ID}$: 
  Upon receiving strings $W_A$ and $W_B$ from user Alice and from
  server Bob, $\F_\ID$ outputs the bit $Y = (W_A \eqq W_B)$ to Bob. In case
  Alice is dishonest, she may choose $W_A = \,\perp$ (which never agrees with honest Bob's input), and (for any choice of $W_A$) the bit $Y$ is also output to Alice.   \vspace{-1ex}
\end{framed}
\vspace{-2ex}
\small
\caption{The Ideal Password-Based Identification Functionality.}\label{fig:identfuncsimpler}
\vspace{-1ex}
\end{figure}

The definition is in the framework described in
Section~\ref{sec:SecDef}; thus, it considers a single execution of the
protocol with an arbitrary distribution for the honest players inputs
and with no input for dishonest players, and security is defined by
information-theoretic conditions on the resulting output states. For
consistency with the above notation (and the notation used
in~\cite{DFSS07}), Alice and Bob's inputs are denoted by $W_A$ and
$W_B$, respectively, rather than $U$ and $V$. Furthermore, note that
honest Alice's output $X$ is empty: $X = \emptyset$.

\begin{definition}[Secure Identification] \label{def:identsimpler}
A password-based quantum identification scheme is {\em $\eps$-secure} if the following properties hold. 
\begin{description}
\item[{\em Correctness:}] For honest user Alice and honest server Bob,
  Bob learns whether their input is equal, except with probability~$\eps$.
\item[{\em Security for Alice:}] For any dishonest server $\dB \in
  \mathfrak{B}$, and for any distribution of $W_A$, the resulting
  common output state $\rho_{W_A Y'}$ (augmented with $W_A$) is such
  that there exists a classical $W'$ that is independent of $W_A$ and
  such that
$$
\rho_{W_A W' Y'|W_A \neq W'} \close{\eps} \rho_{W_A\leftrightarrow
  W'\leftrightarrow  Y'|W_A \neq W'}
\enspace , 
$$
\item[{\em Security for Bob:}] 
For any dishonest user $\dA \in \mathfrak{A}$, and for any distribution of $W_B$, the resulting common output state $\rho_{W_B Y X'}$ (augmented with $W_B$)
is such that there exists a classical $W'$ (possibly $\perp$) independent of $W_B$, such
that if $W_B \neq W'$ then $Y = 1$ with probability at most $\eps$,
and \emph{if $W_B = W'$, Bob's output is $Y=1$}. Furthermore, we have that
$$
\rho_{W_B W' X'|W' \neq W_B} \close{\eps} \rho_{W_B\leftrightarrow W'\leftrightarrow X'| W'\neq W_B} \enspace .
$$ 
\end{description}
\end{definition}
The only difference to Definition~\ref{def:ident} from~\cite{DFSS07}
is that we additionally require for the security for Bob, that he
accepts in case that $W_B = W'$. This small change allows us to
achieve a more natural functionality compared to the case where we leave
undefined what happens in case $W_B = W'$. We note that the protocol
proposed in~\cite{DFSS07} fulfills also this strengthened
Definition~\ref{def:identsimpler}. In Step~5 of their protocol, if
dishonest Alice sends a string $Z$ which is inconsistent with any of
the possible strings $S_j$ corresponding to Bob's passwords, $W'$ is
set to $\perp$. This $W'$ is independent of $W_B$ and as Bob always
rejects, dishonest Alice does not learn any additional
information about $W_B$.

\begin{proposition}
A quantum protocol satisfying Definition~\ref{def:identsimpler}
$\eps$-securely implements the functionality $\F_{\ID}$ from
Figure~\ref{fig:identfuncsimpler} according to
Definition~\ref{def:security}. 
\end{proposition}

\begin{proof}
Correctness follows immediately.

{\em Security for Alice:} as in Proposition~\ref{prop:ident}.

{\em Security for Bob:} 
Consider $W'$ which is guaranteed to exist by
Definition~\ref{def:identsimpler}.  Let $U = W'$ and $X = (W' \eqq W_B)$.
Recall that by the requirement of Definition~\ref{def:identsimpler}, $W'$ is
independent of Bob's input $W_B$.  

Since by Definition~\ref{def:ident} the probability for Bob to decide
that the inputs are equal, $Y=1$, does not exceed $\eps$ if $W_B \neq
W'$, and Bob accepts, $Y=1$, if $W_B = W'$, we have that
\begin{align*}
P_{W' W_B X Y} &= \Pr[W_B = W'] \cdot P_{W' W_B X Y|W_B = W'} + \Pr[W_B
\neq W'] \cdot P_{W' W_B X Y|W_B \neq W'}\\
&= \Pr[W_B = W'] \cdot P_{W' W_B \F_\text{\ID}(W',W_B)|W_B = W'} +
\Pr[W_B \neq W'] \cdot P_{W' W_B X Y|W_B \neq W'}\\
&\approx_\eps \Pr[W_B = W'] \cdot P_{W' W_B \F_\text{\ID}(W',W_B)|W_B =
  W'} + \Pr[W_B \neq W'] \cdot P_{W' W_B \F_\text{\ID}(W',W_B)|W_B \neq W'}\\
&= P_{W' W_B \F_\text{\ID}(W',W_B)}
\end{align*}

Finally, we have 
\begin{align*}
\rho_{W_B Y U X X'} &= 
\Pr[W_B \neq W'] \cdot \rho_{W_B Y W' X X'|W_B
\neq W'} + \Pr[W_B = W'] \cdot \rho_{W_B Y W' X X'|W_B = W'}
\end{align*}
In the case $W_B=W'$, we have by construction that $X=Y=1$ and
therefore, we obtain that $\rho_{W_B Y W' X X'|W_B = W'} =
\rho_{\MC{W_B Y}{W' X}{ X'}|W_B = W'}$. If $W_B \neq W'$, it follows
from Definition~\ref{def:identsimpler} that $\rho_{W_B W' X'|W' \neq
  W_B} \close{\eps} \rho_{W_B\leftrightarrow W'\leftrightarrow X'|
  W'\neq W_B}$. Furthermore, the bit $X$ is fixed to $0$ in case $W_B
\neq W'$ and we only make an error of at most $\eps$ assuming that
Bob's output $Y$ is always $0$ and therefore,
\begin{align*}
\rho_{W_B Y W' X X'|W_B \neq W'} &\approx_\eps \rho_{W_B (Y=0) W'
  (X=0) X'|W_B \neq W'} \\
&\approx_\eps \rho_{\MC{W_B (Y=0)}{W' (X=0)}{X'}|W_B \neq W'}\\
&\approx_\eps \rho_{\MC{W_B Y}{W' X}{X'}|W_B \neq W'}
\end{align*}
Putting things together, we obtain
\begin{align*}
\rho&_{W_B Y U X X'}\\ &\epsclose[3\eps]
\Pr[W_B \neq W'] \cdot \rho_{\MC{W_B Y}{W' X}{X'}|W_B \neq W'} 
+ \Pr[W_B = W'] \cdot \rho_{\MC{W_B Y}{W' X}{ X'}|W_B = W'} \\
&= \rho_{\MC{W_B Y}{W' X}{X'}} \, ,
\end{align*}
where we used Lemma~\ref{lemma:right->mid} and~\ref{lemma:event} in the last step.

\vspace{-3.5ex}
\qed
\end{proof}

\end{document}